\UseRawInputEncoding
\documentclass[twocolumn,aps,pra,bibfootnote,preprintnumbers, mathtools, superscriptaddress, amsmath,amssymb]{revtex4-1}

\usepackage{graphicx}
\usepackage{color}
\usepackage{hyperref}
\usepackage{braket}
\usepackage{orcidlink}

\newcommand{\ms}[1]{\mbox{\scriptsize #1}}

\begin{document}

\title{Quantum open system identification via global optimization: \\ optimally accurate Markovian models of open systems from time-series data }

\author{Zakhar Popovych
\orcidlink{0009-0005-1641-2653}}
\email{zpopovych@tulane.edu}
\affiliation{Department of Physics, Tulane University,  New Orleans, LA 70118, USA}

\author{Kurt Jacobs \orcidlink{0000-0003-0828-6421}}
\email{dr.kurt.jacobs@gmail.com} 
\affiliation{DEVCOM Army Research Laboratory, Adelphi, Maryland 20783, USA}
\affiliation{Department of Physics, University of Massachusetts at Boston, Boston, Massachusetts 02125, USA}
\author{Georgios Korpas}
\email{georgios.korpas@hsbc.com} 
\affiliation{HSBC Lab, Innovation \& Ventures, 8 Canada Square, London E14 5HQ, U.K.}
\affiliation{%
Department of Computer Science, 
Czech Technical University in Prague, Karlovo nam. 13, Prague 2, Czech Republic}
\affiliation{Archimedes Research Unit on AI, Data Science and Algorithms, Athena Research and Innovation Center, Maroussi, Greece}
\author{Jakub Marecek}
\email{jakub.marecek@fel.cvut.cz}
\affiliation{%
Department of Computer Science, 
Czech Technical University in Prague, Karlovo nam. 13, Prague 2, Czech Republic}

\author{Denys I. Bondar \orcidlink{0000-0002-3626-4804}}
\affiliation{Department of Physics, Tulane University,  New Orleans, LA 70118, USA}
\email{dbondar@tulane.edu}

\date{\today}
\begin{abstract}
Accurate models of the dynamics of quantum circuits are essential for optimizing and advancing quantum devices. Since first-principles models of environmental noise and dissipation in real quantum systems are often unavailable, deriving accurate models from measured time-series data is critical. However, identifying open quantum systems poses significant challenges: powerful methods from systems engineering can perform poorly beyond weak damping (as we show) because they fail to incorporate essential constraints required for quantum evolution (e.g.,\ positivity). Common methods that can include these constraints are typically multi-step, fitting linear models to physically grounded master equations, often resulting in non-convex functions in which local optimization algorithms get stuck in local extrema (as we show). In this work, we solve these problems by formulating quantum system identification directly from data as a polynomial optimization problem, enabling the use of recently developed global optimization methods. These methods are essentially guaranteed to reach global optima, allowing us for the first time to efficiently obtain the \textit{most accurate Markovian model} for a given system. In addition to its practical importance, this allows us to take the error of these Markovian models as an alternative (operational) measure of the non-Markovianity of a system. We test our method  with the spin-boson model --- a two-level system coupled to a bath of harmonic oscillators --- for which we obtain the exact evolution using matrix-product-state techniques. We show that polynomial optimization using moment/sum-of-squares approaches significantly outperforms traditional 
optimization algorithms, and we show that even for strong damping Lindblad-form master equations can provide accurate models of the spin-boson system (fractional errors at the level of $10^{-3}$). Using Breuer's measure of non-Markovianity as an example we show that there can be large differences between such measures and the accuracy of the best Markovian models, and both measures indicate that the Markovianity of the spin-boson model is highly non-monotonic with the damping rate. We expect that our method can be extended to non-Markovian models and has the potential to significantly improve quantum system identification and control, with applications ranging from quantum gate design to error correction.  

\end{abstract}
\maketitle

\section{Introduction \label{sec:intro}}

The precise control of multi-component quantum systems holds great promise for the development of new technologies with applications in information processing~\cite{nielsen_quantum_2010, obrien_photonic_2009,huggins_unbiasing_2022} and precision sensing~\cite{yu_capability_2021, xie_biocompatible_2022, marciniak_optimal_2022}. 
However, quantum systems are extremely sensitive to their environment, which introduces noise, dissipation, and decoherence. Given the complexity of these environmental interactions, first-principles models are not available for many systems, making it increasingly important to derive accurate models from time-series measurements.  

The evolution of isolated quantum systems is linear. Powerful methods have been developed in statistics, control, and system engineering to extract models of linear dynamical systems from time-series measurements, a subject referred to as \textit{system identification}, often abbreviated using the acronym SID~\cite{ljung_system_1987}. These methods have proved useful for quantum systems~\cite{mccauley_accurate_2020} but they have a significant limitation: it is not possible to impose constraints on the norm and positivity of the quantum density matrix, which are essential properties. We show that while linear SID methods can work very well for weak damping (although they lack the physical interpretation that comes with master equation models) they can perform very badly for stronger damping by losing positivity and stability. The only way found to date to perform system identification while preserving these properties is to use numerical optimization to fit the data to general models for which these properties are built in~\cite{zhang_identification_2015, samach_lindblad_2022, xue_gradient_2021, mazza_machine_2021}. Numerical optimization is problematic because of the existence of complex optimization landscapes with multiple local minima. Achieving a global optimum typically requires numerous repetitions with randomized initial conditions. Simulated annealing was developed to help solve this problem, but it still requires searching large areas of the landscape. 

Significant recent advances in optimization theory have resulted in methods that can deterministically obtain the global optimum for any polynomial function~\cite{henrion_moment-sos_2021}. Here, we show that system identification can be formulated as the optimization of a polynomial function. This is a significant step forward in system identification that produces models with guaranteed properties, because it is able to find the \text{best possible} model given the dynamical time series. Admittedly the use of polynomial optimization via the moment/sum-of-squares algorithm\cite{parrilo2000structured, lasserre_global_2001, nie_moment_2023} is still relatively new as a practical tool, and our results indicate that current implementations will require further refinements; we find one situation in which the sparsity-exploiting algorithm TSSOS \cite{wang2019tssos} oddly fails to find global optima. Nevertheless, we were able to address this issue with the appropriate choices of master equation ansatz, and all our results are consistent with our method obtaining global optimal solutions: testing the method against all the optimization algorithms in the NLOpt library, the polynomial optimization obtains models that always meet and often significantly exceed the accuracy of those obtained with the other, gradient-based, optimization algorithms. 


The optimization method that we employ to solve the polynomial quantum open system identification problem is called the moment/sum-of-squares hierarchy~\cite{parrilo2000structured, henrion_moment-sos_2021, nie_moment_2023}. While the method is non-trivial to implement, libraries have already been written (e.g., in Julia ~\cite{wang_tssos_2020}, Matlab \cite{lofberg2004yalmip}, and C++ \cite{heller2016gposolver}) making it straightforward to use.

While we consider here only the identification of the best \textit{Markovian models} for open quantum systems, we expect that non-Markovian models can also be readily cast as polynomial optimization problems in a similar way. 

To test our method on a non-trivial and non-Markovian system we apply it to time-series data obtained from exact simulations of the spin-boson model, in which a two-level system is coupled to a bath of harmonic oscillators. We employ the standard Ohmic bath and perform the exact simulations using the matrix-product-state method introduced by Bulla~\textit{et al.}~\cite{bulla_numerical_2003} and refined by Chin~\textit{et al.}~\cite{chin_exact_2010,prior_efficient_2010}. While the spin-boson model generally becomes more non-Markovian the larger the damping, we find that its non-Markovianity is far from being monotonic with the damping rate. 

We compare the performance of our method both to a similar method obtained by replacing polynomial optimization with state-of-the-art traditional optimization methods (as discussed above), and to previous methods that employ traditional linear system identification, including exact and SVD-based dynamic mode decomposition (DMD)~\cite{kutz_dynamic_2016, goldschmidt_data-driven_2022}. We will describe these methods below. For sufficiently weak damping we show that traditional SID methods for linear systems, when used without the additional step of attempting to fit to master equation forms, can be very accurate although they lack the additional benefit of a physical interpretation and guaranteed properties.  

This paper is organized as follows. In Sec.~\ref{sec:sid_pop_theory} we construct our method for system identification, which employs polynomial optimization to find the best model from a general parameterized Markovian master equation. We initially set this up using the most general form for the master equation due Kossakowski, which has the important property that a given master equation is defined by a unique set of parameters. Neverthless, any form of the master equation can be used, and we find that more restricted master equations can be better for some problems. In Section~\ref{sec:applications} we apply the method to identify Markovian models for a non-trivial spin-boson system across a wide range of coupling strengths (damping rates). In Sec.~\ref{sec:discussion}, we discuss in some detail how traditional methods for linear system identification work and compare the performance of our method to previous methods that employ them.  Finally, in Sec.~\ref{sec:summary}, we present our conclusions regarding our method and discuss considerations for its further development and applications.

\section{System Identification using Markovian models}\label{sec:sid_pop_theory}

In system identification, the natural input consists of sampled observations of the system's state taken at discrete time intervals. 
For quantum systems, the density matrix $\rho$ serves as input, however, the elements of the it are not directly accessible through measurements. To reconstruct the density matrix at a specific time - that is to perform quantum state tomography which amounts to inverting the mapping between the measurement results and the elements of the density matrix \cite[Chapter 8]{nielsen_quantum_2010} - one must perform a series of experiments that involve repeatedly preparing the system in the same initial state, allowing it to evolve under identical conditions, and measuring different observables on each iteration. By statistically analyzing the outcomes of these measurements across the ensemble, one can infer the elements of the density matrix and thus fully characterize the state of the system at that time.


It is therefore natural to break the problem of quantum system identification into two parts. In the first part the density matrix of the system is determined at a sequence of times using whatever measurement and tomography procedure is most appropriate for the system. In the second part the sequence of density matrices is used to determine the dynamics of the system. We are concerned here solely with the second part of this procedure, assuming we have sufficient knowledge of the output states. Thus, term ``system identification'' refers  to the process of determining the dynamics based on the observations from a sequence of states of the system, not to the process by which this sequence of states is obtained. 

Let $\{ t_j \}_{j=1}^N$ denote a sequence of time points at which the system's state is sampled. At each time $t_j$, the corresponding state of the system is represented by the density matrix $\rho_j = \rho(t_j) \in \mathcal{H}$, where $\mathcal{H}$ is the Hilbert space of the system. The elapsed time between consecutive samples is defined as $\Delta t_j = t_j - t_{j-1}$ for $j = 2, \ldots, N$. To reconstruct each $\rho_j$, we perform quantum state tomography at time $t_j$, which entails preparing the system in a consistent initial state $\rho_0$, allowing it to evolve under the system dynamics for a duration $t_j$, and performing a set of measurements sufficient to determine the elements of $\rho_j$. This procedure must be repeated multiple times to obtain statistically significant data for accurate state reconstruction. The collection $\{ \rho_j \}_{j=1}^N$ thus provides a discrete-time representation of the system's evolution, sampled at times $\{ t_j \}_{j=1}^N$.


We consider first an SID procedure that allows to directly obtain a Markovian master equation. Identifying a master equation directly requires that the time intervals between consecutive states, $\rho_j = \rho(t_j)$ and $\rho_{j+1}=\rho(t_{j+1})$, are small enough that a low-order approximation to the evolution gives the change in the state, $\rho_{j+1} - \rho_j$, with sufficient accuracy. We show below how to relax this requirement by employing a two-step process. Before we describe the system identification process, it is important to consider how to represent the master equation. 

\subsection{Unique representation of  Markovian evolution}

The Gorini-Kossakowski-Sudarshan-Lindblad master equation~\cite{gorini_completely_1976, lindblad_generators_1976} (sometimes called the GKSL or simply the Lindblad equation) is sufficient to describe all quantum systems whose environment acts in a \textit{Markovian} manner~\cite{li_concepts_2018}. 
Recall that a ``Markovian" environment implies that the system's evolution at time $t$ depends exclusively on its state $\rho(t)$ at that instant, without any dependence on its prior states $\rho(t')$ for $t' < t$, i.e., there no memory effects arise from system-environment interactions. This condition necessitates that any perturbations the system imparts to the environment do not influence the environment's subsequent interactions with the system.

GKSL or Lindblad master equations are widely used to model open quantum systems and reconstruct quantum processes, including in solid-state qubits~\cite{howard_quantum_2006, samach_lindblad_2022, zhang_non-markovian_2021} and trapped-ion qubits~\cite{ben_av_direct_2020, dobrynin_compressed-sensing_2024}—a procedure generally referred to as Lindblad quantum tomography.

The most common Lindblad form of the  GKSL equation can be written as~\cite{lindblad_generators_1976, jacobs_quantum_2014}:
\begin{align}
    \frac{d\rho}{dt} 
    &=  \mathcal{L}_{\{J_k\}, H}(\rho) \notag\\
    &= - i[H, \rho] + \frac{1}{2} \sum_{k}\left( 2 J_k \rho J_k^\dagger - J_k^\dagger J_k \rho - \rho J_k^\dagger J_k \right). \label{eq:lindblad}
\end{align}
Here, the Hamiltonian $H$ is a Hermitian operator that governs the unitary evolution of the system, and the Lindblad jump operators $\{J_k\}$ are arbitrary and describe the dissipative evolution induced by the environment.

While the Lindblad form of the Markovian master equation \eqref{eq:lindblad} is simple and convenient for modeling, it is less suitable for system identification due to the non-uniqueness of the operators $H$ and $\{J_k\}$. This non-uniqueness means that there can be infinitely many sets of operators $\{H, \{J_k\}\}$ that yield the identical dynamics. Therefore, if we pose an optimization problem to find the operators that best fit the data, we encounter the issue of a continuum of solutions. Such a behavior of the objective function can be problematic for many minimization algorithms. It would be preferable to have a representation for the master equation in which there is a one-to-one mapping between the set of parameters and the resulting dynamics. In fact, the Gorini-Kossakowski-Sudarshan (GKS) \cite{gorini_completely_1976} form of the GKSL master equation, though slightly more complex, is written in terms of parameters that uniquely specify the master equation (the Markovian dynamics); see \cite{yoshida2024uniqueness} for a modern proof.  Concretely, the GKS master equation takes the following form:
\begin{align}
    \frac{d \rho}{dt}
    &= \mathcal{K}_{C, H} (\rho) \notag\\
    &= -i [H, \rho] + \frac{1}{2} \sum_{k,l=1}^{n^2-1} C_{kl} \left([f_k, \rho f_l^\dagger] + [f_k \rho, f_l^\dagger]\right), \label{eq:GKS}
\end{align}
where \( \operatorname{Tr} H = 0 \), \( \operatorname{Tr} f_k = 0 \), \( \operatorname{Tr}(f_k f_j^\dagger) = \delta_{kl}, \quad k, l = 1, 2, \dots, n^2 - 1 \) for \( n \) energy levels in the system, and \( C = (C_{kl}) \) is a complex positive matrix, referred to as the Kossakowski matrix. Note that operators $\{f_k\}$ form the basis of operators for the system, and the above master equation describes all possible valid Markovian evolutions for the systems (all possible Markovian master equations).  

In the master equation \eqref{eq:GKS}, once the set of basis operators $\{f_k\}$ is fixed, the traceless Hamiltonian $H$ and the matrix $C$ are uniquely determined. Therefore, if the system exhibits Markovian behavior, then there is a unique choice of $H$ and $C$ that generates that dynamics. Our system identification task is to determine those parameters (for a Markovian system) and for a non-Markovian system determine the parameters that most closely reproduce the measured dynamics. 

\subsection{Formulating system identification as polynomial optimization}

If $\rho$ is represented as a vector rather than a matrix, then the superoperator $\mathcal{K}_{C, H}$ becomes a matrix. The solution of the master equation~\eqref{eq:GKS} can then be written as:
\begin{align}
    \rho(t) = e^{\mathcal{K}_{C, H}t} \rho(0).
\end{align}

Our goal is to determine the superoperator $\mathcal{K}_{C, H}$ that best approximates the change in the state of the system from each time $t_j$ to the next. That is, we want to find the $\mathcal{K}_{C,H}$ that minimizes the difference between $\exp\left(\mathcal{K}_{C, H} 
\Delta t\right)\rho_{j}$ and $\rho_{j+1}$ when summed over all $j$. This means that we want to minimize the loss function 
\begin{align}
    L(C, H) = \sum_j \bigl\|\rho_{j+1} - \exp\left( \mathcal{K}_{C, H}
\Delta t\right) \rho_j \bigr\|_{p}^\ell, \label{eq:loss_gen}
\end{align}
where $|\ldots |_{p}$ is some suitable $p-$norm and $\ell \in \mathbb{N}$. However, we need to determine $L$ in terms of the elements of $\mathcal{K}$ rather than $e^{\mathcal{K}t}$. So long as the time steps $\Delta t$ are sufficiently small, we can achieve this as follows. We can integrate the master equation~\eqref{eq:GKS} over the time interval $[t_{j-2}, t_{j}]$ to obtain 
\begin{align}
    \rho_j - \rho_{j-2} =&  \int^{t_j}_{t_{j-2}} \frac{d\rho(t)}{dt} dt =  \int^{t_j}_{t_{j-2}} \mathcal{K}_{C, H}[\rho(t)] dt \notag\\
    =& \mathcal{K}_{C, H} \left[\int^{t_j}_{t_{j-2}}\rho(t)dt \right].
\end{align}
To ensure that the objective is polynomial, we start by rewriting the loss function \eqref{eq:loss_gen} with a double step and using the Frobenius norm. The result is 
\begin{align} 
 L(H, C)  = \sum_{j=3}^N 
    \left\|
    \rho_j - \rho_{j-2}
    - \mathcal{K}_{C, H} \left[\int^{t_j}_{t_{j-2}}\rho(t)dt \right] \right\|^2_F  \label{eq:kossak_obj}
\end{align}
We then apply the Simpson approximation method to evaluate the integral:
\begin{align}
    \int^{t_j}_{t_{j-2}}\rho(t)dt = \frac{\Delta t }{3} \left( \rho_{j-2} + 4 \rho_{j-1} + \rho_j \right) +  \mathcal{O}(\Delta t^5). \label{eq:simpson}
\end{align}

To express the loss function as a polynomial in terms of the parameters of the Hamiltonian $H$ and the Kossakowski matrix $C$, we need to choose the set of basis operators $\{f_k\}$ such that $\operatorname{Tr} f_k = 0$ and $\operatorname{Tr}(f_k f_l) = \delta_{kl}$. Finally, we have our system identification problem formulated as a polynomial matrix inequality (PMI) problem \cite{henrion2011inner,kocvara2012pennon} in the unknown parameters of $H$ and $C$: 
\begin{align}
    \underset{C, H}{\text{minimize}} & \sum_{j=3}^N 
    \Bigg\|
    \rho_j - \rho_{j-2} \notag\\
     & \left. \quad - \frac{\Delta t }{3} \mathcal{K}_{C, H} \left(
         \rho_{j-2} + 4 \rho_{j-1} + \rho_j 
    \right)\right\|^2_F, \notag\\
    \text{subject to } & C \succ 0, \notag\\
    & H = H^\dagger, \quad \operatorname{Tr}H = 0.\label{eq:pop_kossak_gen}
\end{align}
While there has been much recent progress \cite[e.g.]{zheng2023sum,guo2024moment} in solving PMIs,
it can be beneficial to reparameterize $C$ with an arbitrary complex matrix $M$ as $C = M^\dagger M$. Similarly, the Hermiticity of Hamiltonian $H = H^\dagger$ can be guaranteed by parameterization $H = X + X^T + i(X-X^T)$, where $X$ is a square matrix with real entries.
In this way, the PMI \eqref{eq:pop_kossak_gen} is reformulated as a polynomial optimization problem (POP). 

Overall, the problem of identifying the master equation from the time series of states $\{  \rho_j\}_{j=1}^N$ can be formulated as the following POP: 
\begin{align} \label{Eq:Loss_gen} 
    \underset{M, H}{\text{minimize}} & \sum_{j=3}^N 
    \Bigg\|
    \rho_j - \rho_{j-2} \\
     & \left. \quad - \frac{\Delta t }{3} \mathcal{K}_{M^\dagger M, H} \left(
         \rho_{j-2} + 4 \rho_{j-1} + \rho_j 
    \right)\right\|^2_F, \notag \\
    \text{subject to} \quad  & H = H^\dagger,\quad \operatorname{Tr}H = 0. \notag
\end{align}
Notice that the constraint $\mbox{Tr}H=0$ allows for unique optimizers in terms of $H$, which aids the recovery of the optimizer. 
Having the Hamiltonian $H$ and Kossakowski matrix $C$, we can readily recover parameters of the master equation in the Kossakowski form \eqref{eq:GKS} and, from that, it is simple to obtain the set of jump operators $J_k$ for the Lindblad form \eqref{eq:lindblad} of the master equation, if desired. 

In solving~\eqref{Eq:Loss_gen}, we can use moment/sum-of-squares approaches, as implemented in TSSOS \cite{wang_tssos_2020}, for example. A caveat is that this implementation is still, to some extent, experimental and does need some more time to develop into a mature technology. Nevertheless,  even in its current stage of development, it outperforms all other optimization methods we have tried, and more importantly, it guarantees that the best model is obtained. 

\section{Application to the spin-boson model\label{sec:applications}}

The spin-boson model is described by the Hamiltonian~\cite{chin_exact_2010, bulla_numerical_2003, leggett_dynamics_1987}
\begin{align}
    H_0 & = \hbar\frac{\nu}{2} \sigma_z + \hbar g \int_0^\Omega \!\! \sigma a^\dagger(\omega) + \sigma^\dagger a(\omega) \, d\omega \nonumber \\ 
    & \;\; + \hbar \int_0^\Omega \!\! \omega a^\dagger(\omega) a(\omega) \, d\omega .
\end{align} 
Here $\sigma_z$ is the Pauli spin-\textit{z} for a two-level (spin-1/2) system and $\sigma$ is the lowering operator for the spin. The operators $a(\omega)$ are annihilation operators for a continuum of modes with frequency $\omega$. Here we have chosen the density of oscillators per unit frequency so that the spectrum is Ohmic~\cite{mccauley_accurate_2020, chin_exact_2010}. The frequency $\Omega$ is referred to as the ``cut-off" frequency of the bath of modes, and is chosen to be much larger than any other timescale of the joint system. The finite value of $\Omega$ introduces some level of non-Markovianity into the dynamics of the spin, as does the non-zero value of $\gamma$ and any deviation from the flatness of spectral density of the oscillators in the bath~\cite{santra_fermis_2017, mccauley_accurate_2020}. This model describes a two-level atom coupled to a one-dimensional electromagnetic field, and the resulting spontaneous emission. When the coupling $g$ is such that $\gamma \equiv g^2/(2\pi) \ll \nu$, where $\nu$ is the transition frequency of the two-level atom, then $\gamma$ gives the decay rate of the atom~\cite{mccauley_accurate_2020}. We will refer to $\gamma$ as the decay rate regardless of the ``quality factor'' of the atom which we can define as $Q \equiv \nu/\gamma$.

We perform essentially exact simulations of the spin-boson model using the fact, detailed in~\cite{chin_exact_2010}, that the coupling to a continuum of oscillators, in which the system is coupled directly to each oscillator, can be exactly mapped into a configuration in which the system is coupled to only a single oscillator that is on the end of a semi-infinite one-dimensional chain of oscillators coupled to each other via nearest neighbors. This allows the chain of the oscillators, which is the bath, to be simulated using a matrix product state method due to Vidal~\cite{vidal_efficient_2003, vidal_efficient_2004}. We use this method to obtain our time-series data of the dynamics of the two-level system for system identification. 

\subsection{Note: the cut-off frequency and the non-Markovian ``initial slip"}

In the absence of driving, all Markovian master equations, being rate equations, exhibit an exponential decay from an initial excited state, so that the initial evolution is approximately a straight line with a slope equal to the decay rate. Conversely, a bath with a cut-off frequency of $\Omega$ will always induce evolution that begins with a slope of zero and curves downward in a manner similar to the cosine function, transitioning to exponential decay after a time of approximately $\Delta t_{\ms{slip}} = 2\pi/\Omega$~\cite{santra_fermis_2017}. This time $\Delta t_{\ms{slip}}$, before the exponential decay of the rate equations ``kicks in" has been referred to as the ``initial slip" in the open system evolution because it appears that the evolution of the system has been delayed by approximately $\Delta t_{\ms{slip}}$~\cite{Davies1974, Gnutzmann96, Gaspard99, Hu92}. This initial slip is a manifestation of non-Markovian behavior due to the finite value of the cut-off frequency $\Omega$. Since we know that the evolution of the two-level system will have this initial slip for all damping rates, we eliminate the error it would contribute to our fitted Markovian models simply by not using the time series during the initial slip period when fitting the model. That is, we are only interested in how well the Markovian models describe the evolution after the initial slip period. Naturally, this period can be made smaller by increasing the value of $\Omega$. 

\subsection{Employing as an ansatz the most general Markovian master equation}

To demonstrate the performance of our SID method on the Ohmic spin-boson model described above we need to chose a master equation \textit{ansatz} to fit to the time-series data. We begin by employing the Kossakowski master equation, described above, which encompases all possible Markovian evolutions of the atom. 

According to Kossakowski's theorem~\cite{kossakowski_general_1973} the Markovian master equation of the form \eqref{eq:GKS} for a two-level system ($n=2$) can be written down using the specific forms of the basis operators $f_k$, traceless Hamiltonian $H$ and Kossakowski matrix $C$. 
The basis operators for a two-level system can be written using the Levi-Civita anti-symmetric tensor,  $\epsilon_{klm}$, and the Kroneker delta,  $\delta_{kl}$:
\begin{align} 
    f_k f_l &= \frac{1}{4}\delta_{kl}I  +\frac{i}{2} \sum_{m=1}^3 \epsilon_{klm} f_m  \notag \\ &\Rightarrow  
    \operatorname{Tr}(f_k f_l)=\frac{1}{2}\delta_{ij}, \quad \operatorname{Tr}(f_k) = 0. \notag 
\end{align}
Thus we can choose the basis operators $f_k$ as the Pauli matrices $\{ \sigma_x, \sigma_y, \sigma_z \}$ scaled by one half: 
\begin{align}
    f_1 = \frac{\sigma_x}{2}, \quad
    f_2 = \frac{\sigma_y}{2},  \quad
    f_3 = \frac{\sigma_z}{2},
\end{align}
where
\begin{align}
    \sigma_x  = 
    \begin{pmatrix}
        0 & 1 \\ 1 & 0
    \end{pmatrix}, \quad
    \sigma_y  = \begin{pmatrix}
        0 & -i \\ i & 0
    \end{pmatrix}, \quad
    \sigma_z = \begin{pmatrix}
        1 & 0 \\ 0 & -1
    \end{pmatrix}. \notag
\end{align}

The Hamiltonian of the two-level system can be expressed as a linear combination of the basis operators with real coefficients $h_k$:
\begin{align}
     H &= \sum_{k=1}^3 h_k f_k
     = \frac{1}{2} 
        \begin{pmatrix} 
            h_3   & h_1 + i h_2 \\
            h_1 - i h_2 & -h_3
        \end{pmatrix}.  \label{eq:H2}
\end{align}
According to the above mentioned Kossakowski theorem \cite{kossakowski_general_1973} the matrix $C$ for the two-level system can be expressed as follows:
\begin{align}\label{eq:C2}
C =  \begin{pmatrix} 
\kappa_1   & - i a_3    &  i a_2 \\
i a_3       & \kappa_2   & -i a_1 \\
-i a_2      & i a_1      & \kappa_3, 
\end{pmatrix}. 
\end{align}
Considering non-unitary dynamics,  in order for the evolution to be completely positive it is necessary and sufficient that:
\begin{equation}
\begin{aligned}
    \kappa_1 + \kappa_2 + \kappa_3 &\geq 0, \\
\kappa_1 \kappa_2  + \kappa_3 \kappa_1  + \kappa_1 \kappa_2  &\geq a_1^2 + a_2^2 + a_3^2 , \\
   \kappa_1 \kappa_2 \kappa_3 &\geq  \kappa_1 a_1^2 + \kappa_2 a_2^2 + \kappa_3 a_3^2.  
\end{aligned}\label{eq:kappAonstr}
\end{equation}

Therefore, using $H$ in the form \eqref{eq:H2} and the Kossakowski matrix $C$ in the form \eqref{eq:C2}, as well as the Simpson approximation \eqref{eq:simpson}, we can express the objective \eqref{eq:kossak_obj}  as a polynomial of nine real variables $\{ h_1, h_2, h_3,\kappa_1,\kappa_2, \kappa_3, a_1, a_2, a_3\}$. Similarly to Eq.~\eqref{eq:pop_kossak_gen} we can cast the optimization problem as:
\begin{align}\label{eq:pop_kossak2}
    \underset{\{h_k, \kappa_k, a_k\}_{k=1}^3}{\text{minimize}} & \sum_{j=3}^N 
    \Bigg\|
    \rho_j - \rho_{j-2} \\
    & \left.
    \quad - \frac{\Delta t }{3}  \mathcal{K}_{C, H} \left(
         \rho_{j-2} + 4 \rho_{j-1} + \rho_j 
    \right)\right\|^2_F, \notag\\
    \text{subject to } & C \text{ has the form~\eqref{eq:C2}}, \notag\\
    & H \text{ has the form~\eqref{eq:H2}}, \notag\\
    & \text{ with constraints  given by ~\eqref{eq:kappAonstr}.} \notag
\end{align}

Now we have a constrained polynomial optimization problem that can be solved easily and naturally with the moment/sum-of-squares hierarchy, e.g., using the TSSOS library~\cite{wang_tssos_2020}. 

\subsection{Results from employing the most general Markovian master equation}

To fit the master equation ansatz to the spin-boson dynamics we used the evolution of four initial states for the atom (the exited state $\ket{1}$, the ground state $\ket{0}$, and the Fourier states $(\ket{0} \pm \ket{1})/\sqrt{2}$). We then tested the accuracy of the resulting Markovian model by comparing the evolution it provided with against the exact evolution of $20$ states of the atom that form the points of a dodecahedron on the Bloch sphere. In fact, due to the symmetry of the initial states used for fitting (optimizing) the model and that of the dodecahedron, this gives us effectively 10 distinct exact trajectories with which to compare. To quantify the error between the model and the exact evolution we used the infidelity, defined as $1-F(\rho_1,\rho_2)$ in which 
\begin{align}
F(\rho_1, \rho_2) = \left( \operatorname{Tr} \sqrt{\sqrt{\rho_1} \rho_2 \sqrt{\rho_1}} \right)^{2}
\label{eq:fidelity}
\end{align} 
is the fidelity~\cite{jozsa_fidelity_1994}. 
Clearly, to obtain an accurate model a sufficient duration of time-series data is required, and we will discuss our findings on that below. First, we simply report the results in which sufficiently long time series were used to obtain the most accurate models (with one exception). 

\begin{figure}[t]
    \includegraphics[width = 1 \columnwidth]{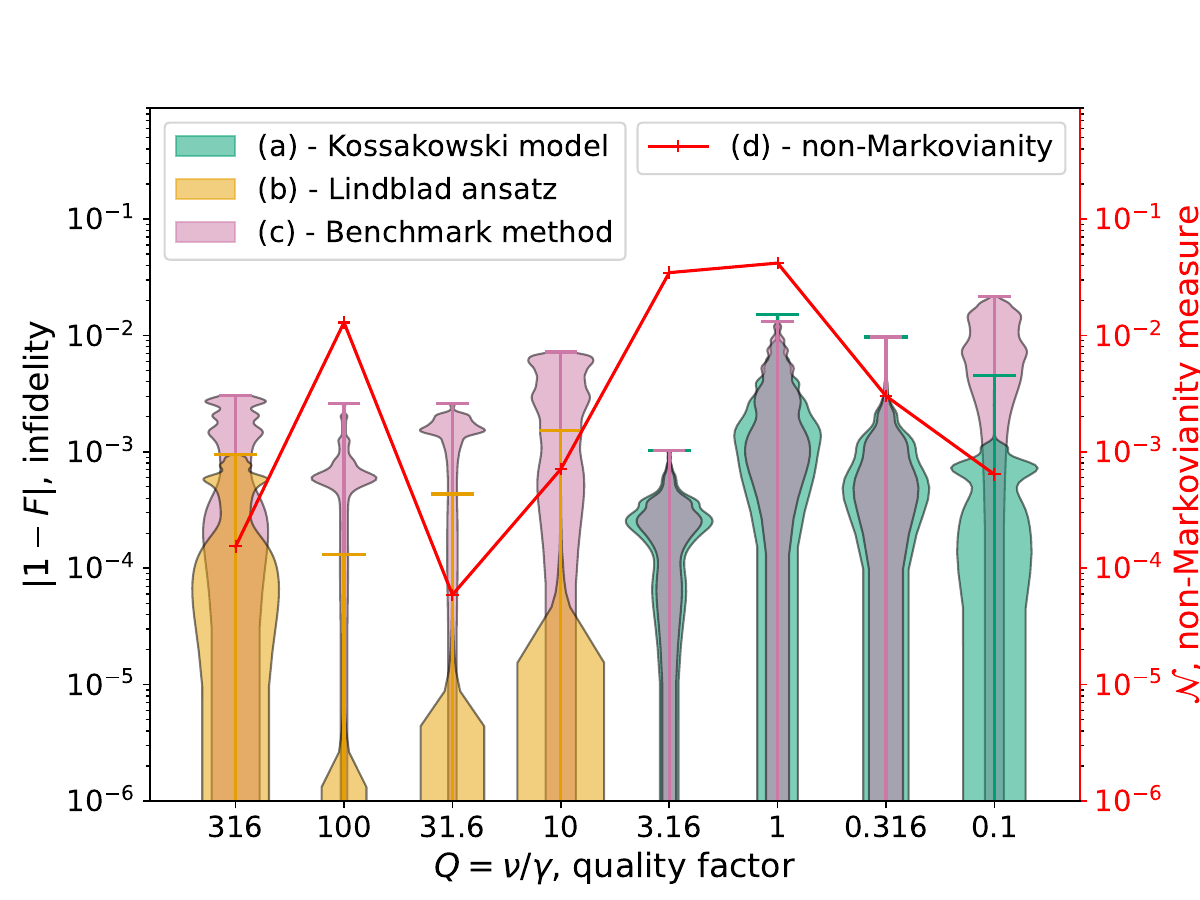}
    \caption{Performance of Markovian models obtained with polynomial (global) optimization ((a) and (b)) compared with a benchmark traditional optimization method from NLOPT (c), and the Breuer measure of non-Markovianity, all for the spin-boson system with a range of $Q$-factors. 
    \textit{Left axis:} Infidelity between the optimized identified model and the true evolution, $1-F$, where the fidelity $F(\rho^{M}, \rho^{SB})$ is defined in Eq.\eqref{eq:fidelity}. The ``violin plots'' give the distribution of infidelities over all data points on the 10 distinct example trajectories (see text) for each of the models obtained with the three different methods. It is an important feature that the polynomial optimization always provide a model at least as good as the traditional search method. The former is essentially guaranteed to obtain the best possible model. (a) Polynomial optimization (POP) using the  Kossakowski master equation, Eq. \eqref{eq:pop_kossak2}, see Ref. \cite{popovych_kossak_pop_2024} regarding a Julia code used; (b) POP using the selected Lindblad ansatz, Eq.\eqref{eq:pop_lindblad2}, see Ref. \cite{popovych_lindblad_pop_2024} regarding a Julia code used;  (c) The best performing  NLOPT optimiser (LD\_SLSQP) for comparison, using the corresponding ansatz, and optimizing the Fidelity explicitly, see Ref. \cite{popovych_kossak_benchmark_2024, popovych_lindblad_benchmark_2024}regarding a Julia code used. 
    \textit{Right axis:} Brauer's non-Markovianity measure $\mathcal{N}$\cite{breuer_measure_2009},  approximated over all available pairs $\{\rho^{(1)}, \rho^{(2)}\}$ of spin-boson system trajectories  using Eq.\eqref{eq:nonmark_breuer}, see Ref. \cite{popovych_non-mark-estims_2024} regarding a Julia code used. See Ref. \cite{popovych_pop_plot_2024} regarding a python code used to generate this figure.
    \label{fig:kossak_vs_lindblad_fids}}
\end{figure}

In Fig.~\ref{fig:kossak_vs_lindblad_fids} we display the results of employing the fully general (and uniquely parameterized) master equation to perform SID for the four largest values of the damping rate $\gamma$ that we explored (these are shown in green). These ``violin plots'' give the distribution of infidelity on all data points for the model obtained for each value of $Q$. The reason we do not show the results of this approach for the four smallest values of $\gamma$ is that it did not converge for those values. It should be emphasized that this failure is, in fact, surprising --- rigorous theory, backed up by an increasing body of numerical results assert that polynomial optimization algorithms find global optima. 
We find that when we replace the general Kossakowski form with a much more restricted master equation polynomial optimization once again performs better that all the other optimization methods we tried. We discuss the results when using the more restricted form in the following. 

\begin{figure}[t]
    \includegraphics[width = 1 \columnwidth]{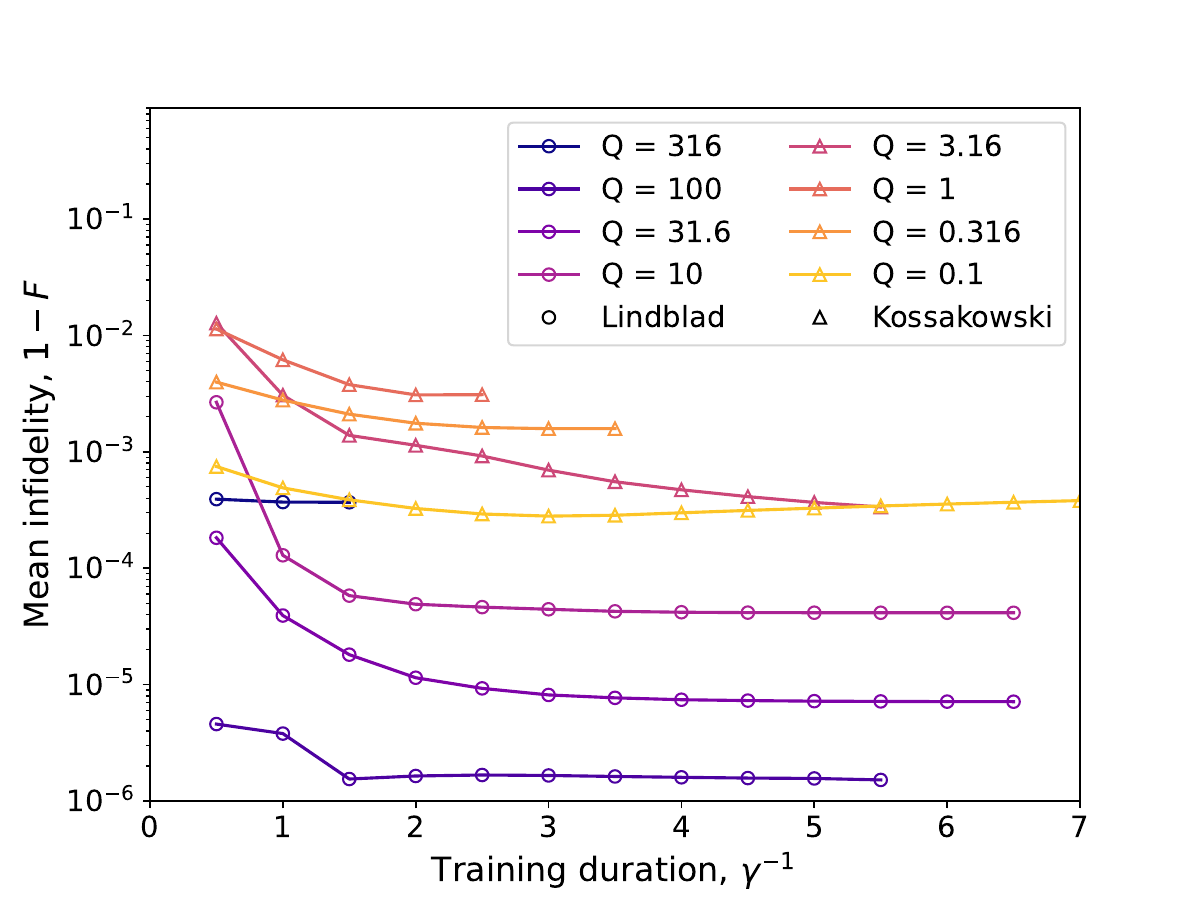}
    \caption{The performance of Markovian master equation models obtained using polynomial (global) optimization as a function of the length of the time series employed, and for eight quality factors, $Q$. We see from this that different quality factors require different durations to obtain the best model. For all the quality factors we investigated, for the simulation times we used, all have reached their optimal models except for $Q=3.16$ which appears to be close. See Ref. \cite{popovych_pop_plot_by_train_2024} regarding a python code used to generate this figure.
\label{fig:train_duration}}
\end{figure}

We also show the accuracy of the models obtained using the best-performing optimization algorithm from the NLOPT optimization toolbox~\cite{steven_g_johnson_nlopt_2007}. The algorithm was SLSQP, which is based on sequential quadratic programming \cite{boggs1995sequential,gill2011sequential}. For this and the other NLOPT optimization algorithms, we also tried using fidelity as a loss function rather than the algebrized (``polynomialized'') loss used for POP, since it is infidelity that we use to characterize the error in the model. The other optimization algorithms either performed poorly or took prohibitively long to run. It is important to emphasize that the moment/sum-of-squares approach either performs significantly better than the other optimization algorithms, or approximately the same. This is a numerical confirmation of the mathematical claim that this optimization method indeed finds the best possible Markovian models because it obtains a global optimum. The importance of this feature cannot be over-stated, and this is the first time that finding the best model has been possible. 

Not only is finding the best model of practical importance, but it also means that we can use the error of this model as an operational measure of the non-Markovianity of the spin-boson evolution. It is interesting therefore to compare this with another measure of non-Markovianity. We choose Breuer's measure, the definition of which we give in Appendix~\ref{AppA}. We plot Breuer's measure in Fig.\ref{fig:kossak_vs_lindblad_fids} for the different $Q$-factors in red. For the four smallest $Q$ values (the largest damping rates) there is generally, although not completely, agreement between the two measures.  

From Fig.\ref{fig:kossak_vs_lindblad_fids}, we see that for $Q = 3.16$ and $Q = 0.1$ the best Markovian master equation provides a pretty good model of the spin-boson system --- it achieves errors almost always below $10^{-3}$.  


The Julia code that implements the SID method using the general Kossakowski form of the master equation is available on GitHub~\footnote{\url{https://github.com/zpopovych/quantum-open-systems-polynomial-sid/blob/main/01_Kossakowski_CONSTR_train-test.ipynb}}. 
The parameters of the identified Markovian master equations in Kossakowski form are stored in \texttt{HDF5} format and also available on GitHub~\footnote{\url{https://github.com/zpopovych/quantum-open-systems-polynomial-sid/blob/main/KOSSAK_CONSTR_TSSOS_treshold_1e-15_FROB_QO_2024-Sep-06_at_11-57.h5} (Note: The file consists of the groups that correspond to the damping rates $\gamma$. In each group you can find the Kossakowski matrix in dataset \texttt{"C"} and Hamiltonian in dataset \texttt{"H"})}. 



\subsection{Employing a simpler Lindblad-form ansatz for higher $Q$}

We find that we can obtain very accurate models for weaker damping by using a fairly simple form of the Lindblad master equation. Polynomial optimization performs much better than any other optimization algorithm in our trial set, indicating that it is reaching the global optimum. Note that as the damping gets weaker we expect that the dynamics is more Markovian and thus master equations should provide increasingly accurate models. 

In the Lindblad form of the master equation, the ``jump operators'' or ``Lindblad'' operators often have simple interpretations as to their physical meaning. A pure damping process is given by the Lindblad operator 
\begin{align}
    J_1 = \begin{pmatrix} 0 & 0 \\ r & 0 
   \end{pmatrix} = r \sigma, \label{eq:j1}
\end{align}
Dephasing processes in each of the $x$, $y$, and $z$ bases are given respectively by:
\begin{align}
J_2 = p \sigma_x, \,
J_3 = p \sigma_y, \,
J_4 = p \sigma_z. \label{eq:j234}
\end{align}
We find that including these four decoherence processes in our Lindblad ansatz is sufficient to obtain accurate models. In addition, we find that we can set all the dephasing processes to the same rate, $p$, so that the irreversible part of the master equation is characterized by only two real parameters, $r$ and $p$. 

Using the general form of $H$ given in Eq. \eqref{eq:H2} and the four Lindblad operators above, the loss function is a polynomial in only five real variables $\{ h_1, h_2, h_3, r, p\}$. The polynomial optimization problem reads: 
\begin{align} \label{eq:pop_lindblad2}
    \underset{ h_1, h_2, h_3, r, p,}{\text{minimize}} & \sum_{j=3}^N 
    \Bigg\|
    \rho_j - \rho_{j-2} \\
    & \left.
    - \frac{\Delta t }{3}  \mathcal{L}_{J_1, \ldots, J_4, H} \left(
         \rho_{j-2} + 4 \rho_{j-1} + \rho_j 
    \right)\right\|^2_F, \notag\\
    \text{subject to } & J_1 \text{ is of the form~\eqref{eq:j1}}, \notag\\
    & J_2, \ldots, J_4 \text{ is of the form~\eqref{eq:j234}}, \notag\\
    & H \text{ is of the form~\eqref{eq:H2}}. \notag
\end{align}

In Fig.\ref{fig:kossak_vs_lindblad_fids} we show the results of our POP system identification (model optimization) in yellow for the four largest values of $Q$.  Clearly, POP obtains much better models than SLSQP for all four values, and for $Q=100$ the best model is usually accurate to well below $10^{-4}$. This is, in fact, the first particularly surprising result we have obtained. Note that the dynamics of an open quantum system is only expected to be truly Markovian in the limit of high $Q$ and high cut-off frequency, $\Omega$~\cite{santra_fermis_2017}. Naively, then, we would expect that for a Markovian model to fit the evolution with errors of $10^{-3}$ it would require $Q \sim 10^{3}$ and $\Omega/\gamma \sim 10^{3}$. (For our spin-boson system we use $\Omega = 10\nu$ so $\Omega/\gamma$ is always equal to $10Q$.) Instead, we find that for $Q = 100$ the Markovian model is accurate to $\sim 10^{-5}$. 

An even greater surprise is that when we increase $Q$ to 316 the accuracy of the best Markovian model \textit{drops} from $10^{-5}$ to $10^{-3}$. We were so surprised by this result that we spent considerable time ensuring that our exact simulations of the spin-boson system were sufficiently accurate and that we were using a sufficient density of data points. 

It is especially interesting to compare the accuracy of the Lindblad models to Breuer's measure for the higher $Q$ values. We find that there is a big difference between Breuer's measure and the accuracy of the models, indicating that non-Markovianity has more than one independent manifestation~\cite{Li_2018}. The big difference in model accuracy, and thus implied non-Markovianity, between the two highest $Q$ values is also seen in Breuer's measure, but in the opposite direction. This provides some confirmation that measures of non-Makovianity can change in unexpected ways, giving us additional confidence that the surprising change we see in the model accuracy is not due to errors in our simulations. 

Finally, it is important to note that the SLSQP optimization method, the best performing of the optimization algorithms in the NLOPT library, only obtains the globally optimal Markovian model in three out of the eight values of $Q$ that we explore. This confirms that the optimization problem for model fitting is a very difficult one, containing a large number of non-optimal extrema. 

Julia code for the simple Lindblad ansatz is available on GitHub~\footnote{\url{https://github.com/zpopovych/quantum-open-systems-polynomial-sid/blob/main/01_Lindblad_CONSTR_train-test.ipynb}}.
Parameters of the identified Markovian master equations in Lindblad form are stored in \texttt{HDF5} format and available on GitHub~\footnote{\url{https://github.com/zpopovych/quantum-open-systems-polynomial-sid/blob/main/LINDBLAD4_CONSTR_TSSOS_treshold_1e-9_FROB_QO_2024-Sep-06_at_16-19.h5}}. The file consists of the groups that correspond to the damping rates $\gamma = \nu/Q$. In each group you can find the identified Hamiltonian in dataset \texttt{"H"} and the four jump operators in datasets  \texttt{"J1"}, \texttt{"J2"}, \texttt{"J3"} and \texttt{"J4"}.

\section{Comparison with traditional linear system identification  \label{sec:discussion}}

The identification of quantum systems has been a long-standing topic of interest in the research community~\cite{mabuchi_dynamical_1996}. Traditionally, many approaches focused on adapting well-established linear system identification techniques from classical control theory to quantum systems, followed by fitting the parameters of the identified models to satisfy quantum-mechanical constraints. However, this often led to a trade-off between model accuracy and physical consistency.

Standard linear system identification methods developed in other fields can readily be applied to the identification of unitary evolution in closed quantum systems. For example, Goldschmidt \cite{goldschmidt_data-driven_2022} employed  Dynamic Mode Decomposition (DMD) (see below) \cite{budisic_applied_2012, kutz_dynamic_2016}. Zhang and Sarovar \cite{zhang_quantum_2014} previously utilized the eigensystem realization algorithm (ERA) (see below) \cite{juang_eigensystem_1985, juang_applied_1994} as a preliminary step in Hamiltonian identification. More recently, the application of physics-informed DMD (pi-DMD) has been explored to identify linear systems with self-adjoint matrices, which correspond to physical models of closed quantum systems governed by the Schrödinger equation \cite{baddoo_physics-informed_2023}.

While imposing conditions of unitary dynamics (as in Hamiltonian identification) can work well \cite{zhang_quantum_2014, goldschmidt_data-driven_2022}, similar methods proposed for open quantum systems do not always produce accurate physical models. The primary motivation for our work here was to address problems with standard linear SID techniques. Because of this, we describe these methods here in some detail so as to highlight their issues and compare them with the method introduced here. In this section, we will compare our method with both  general linear system identification (LSID) to obtain models (which we show works well for sufficiently low damping) and with LSID used as the first step in fitting master equation models (which performs poorly compared to the method we introduce here). 

\subsection{Two standard Linear system identification methods}

Linear system identification is naturally concerned with models that obey a general equation of the form
\begin{align} \frac{d\mathbf{x}(t)} {dt} = M \mathbf{x}(t) \label{eq:linear_system} 
\end{align} 
in which $\mathbf{x}$ is a vector that undergoes dynamics given by the constant  matrix $M$. When applying LSID to a system one usually allows the state of the potential model, $\mathbf{x}$, to have a higher dimension than the state of the system to be modeled, so that the actual state of the system $\mathbf{y}$ (in our case a vectorized version of the density matrix) is given by some linear transformation, $\mathcal{C}$, that maps from a higher to a lower dimensional space : 
\begin{align} \mathbf{y}(t) = \operatorname{vec}[{\rho(t)}] = \mathcal{C} \mathbf{x}(t) \label{eq:observable_state}.
\end{align} 
The matrix $\mathcal{C}$ is thus generally not square. (It might also be the case that the underlying linear model has a smaller dimension than the system.) 

A linear differential equation \eqref{eq:linear_system} may also be written in discrete form for \textit{snapshots} of the state of the system at time $t_1, t_2,... t_N$ as follows:
\begin{align} \label{eq:linear_system_discrete} 
\mathbf{x}_{j+1} = A \mathbf{x}_j, \quad \text{where} \quad 
A =  \exp (M \Delta t) . 
\end{align}

There are two standard methods for determining the equation of motion matrix $A$ from snapshots of the state of the system at a sequence of times. Dynamical mode decomposition (DMD) developed by Schmid and Sesterhenn~\cite{SCHMID_2010} involves taking $N$ snapshots of the system state, column vectors $\mathbf{y}_n$, $n = 1,2,\ldots N$, taken at regular time intervals separated by $\Delta t$, and defining the matrix  
\begin{align}
    Y_k = \left( \mathbf{y}_{k+1} \cdots  \mathbf{y}_{k+N-1} \right)  = \left[\begin{smallmatrix}
| & | & &| \\
\mathbf{y}_{k+1} & \mathbf{y}_{k+2} & \dots &\mathbf{y}_{k+N-1} \\
| & | & &| \\
\end{smallmatrix} \right]
\end{align}
so that if $A$ gives the evolution over tme $\Delta t$, then 
\begin{align}
    Y_1 = A Y_0. 
\end{align}

If we want to obtain the model of the same rank as observable vector $\mathbf{y}_i$ we can just naively use the least squares estimation:
\begin{align}
\hat{A}  =  Y_1 Y_0^{+},
\end{align}
where $Y_0^{+}$ denotes the Moore-Penrose pseudo inverse of $Y_0$. 
If we want to obtain a reduced-order model, we can use the singular value decomposition (SVD). We wont go into the derivation here (see~\cite{SCHMID_2010}) but the estimated $A$ matrix is given by 
\begin{align}
    \bar{A} = Y_1 W \Sigma^{-1} U^{T} 
\end{align}
in which the $U$, $\Sigma$, and $W$ are defined as the three matrices making up the SVD of $Y_0$: 
\begin{align}
    Y_0 = U \Sigma W . 
\end{align}

The second standard method is the eigensystem realization algorithm (ERA) developed by Juang and Pappa~\cite{Juang85, Katayama05}, in which one creates a larger Hankel matrix of snapshots:
\begin{align} \label{eq:Hankel}
\mathcal{H} &=
\left[\begin{smallmatrix}
 \mathbf{y}_1 & \mathbf{y}_2 &\dots &\mathbf{y}_{N/2} \\
 \mathbf{y}_2 & \mathbf{y}_3 &\dots &\mathbf{y}_{N/2+1} \\
 \vdots &	&\ddots & \vdots \\
 \mathbf{y}_{N/2} & \mathbf{y}_{N/2+1}  &\dots &\mathbf{y}_{N} \\
 \end{smallmatrix} \right].
 \end{align}
This time the dynamical matrix that provides the system model is 
 \begin{align} \label{eq:Hankel}
A' = (U \sqrt{\Sigma})^+_{\uparrow}(U \sqrt{\Sigma})_{\downarrow}
\end{align}
where $M_{\uparrow}$ and $M_{\downarrow}$ -- refers to exclusion of $\operatorname{dim}[\mathbf{y}_i]$ the first and last rows of the matrix $M$ respectively, in which $\mathcal{H} = U \Sigma V$ is the SVD for $\mathcal{H}$. Note that while the size of the dynamical system (the size of $\bar{A}$) obtained with the DMD method is the same as that of the state vector that is obtained from the time-series, the size of $A'$ generated with the ERA method may be much larger. 
Often, one can truncate $A'$ so as to eliminate negligible eigenvalues.

In fact, there is a simple way to extend the DMD method to obtain a model that is larger than the original time-series data (as with ERA) by ``stacking'' the state vectors of $\mathbf{y}$ for different initial conditions together. Thus, if we have four time series with four different initial  conditions, $\mathbf{y}^{(0)}, \ldots, \mathbf{y}^{(3)}$ then we take as our state vector at time $t$ the column vector  
\begin{align}
    \mathbf{z}(t) &= \left( \mathbf{y}^{(0)\ms{T}}(t),  \; \mathbf{y}^{(1)\ms{T}}(t), \; \cdots ,\; \mathbf{y}^{(3)\ms{T}}(t) \right)^{\ms{T}} \\ 
    Z&=
\left[\begin{smallmatrix}
| & | & &| \\
\mathbf{z}_{1} & \mathbf{z}_{2} & \dots &\mathbf{z}_{N-1} \\
| & | & &| \\
\end{smallmatrix} \right] =
\left[
\begin{smallmatrix}
\mathbf{y}^{\ket{0}}_1 & \mathbf{y}^{\ket{0}}_2 &   	& \mathbf{y}^{\ket{0}}_{N-1} \\
\mathbf{y}^{\ket{1}}_1 & \mathbf{y}^{\ket{1}}_2 & \dots & \mathbf{y}^{\ket{1}}_{N-1} \\
\mathbf{y}^{\ket{x}}_1 &  \mathbf{y}^{\ket{x}}_2 & 	& \mathbf{y}^{\ket{x}}_{N-1} \\
\mathbf{y}^{\ket{y}}_1 & \mathbf{y}^{\ket{y}}_2 &   	& \mathbf{y}^{\ket{y}}_{N-1} \\
\end{smallmatrix} \right]  
\end{align}
If there are fewer significant eigenvalues than in the original system, one has achieved \textit{model reduction}, i.e., a simpler model than the original system.


Applying this to system identification of the evolution of the density matrix naturally involves ``vectorizing'' the density matrix (stacking its columns one beneath of the other) which we will write as $\mbox{vec}[\rho]$, performing LSID on the evolution of that vector, and then hoping that the resulting model preserves the norm and complete positivity of the density matrix as well as being a stable evolution (all real parts of the eigenvalues of $\bar{A}$ or $A'$ are non-positive). 

Since the model determined by LSID is only well defined up to a similarity transformation of its state, the state-vector of the model obtained need not correspond directly to the density matrix. 
Since engineering treatments often do not give the transformation required to convert the model so that its state vector is that of the original system, we determine this transformation in Appendix~\ref{AppB}.




\begin{figure}
    \includegraphics[width = 1 \columnwidth]{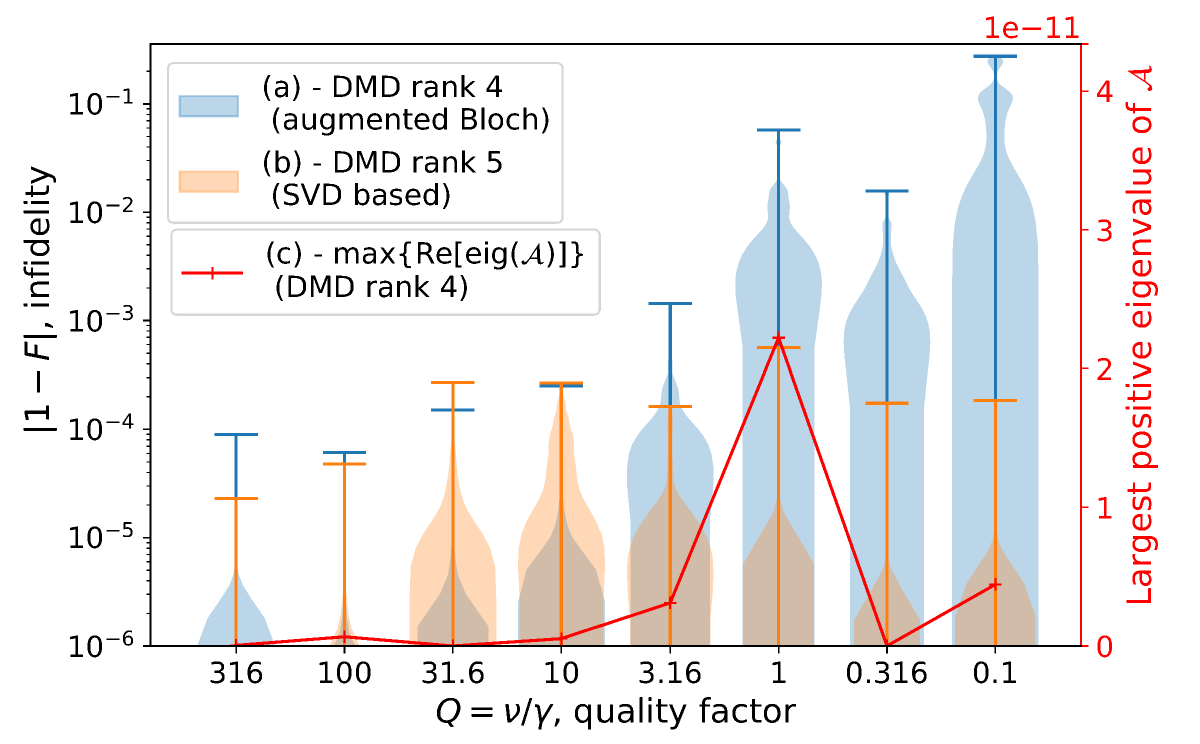}
    \caption{Performance of linear system identification methods for the spin-boson system for different quality factors $Q$. For each value of $Q$ the performance of the model(s) is shown using a ``violin'' plot that shows the distribution of the fidelity of the model over all the data points of the time-series for our 10 distinct initial conditions (half the points of a dodecahedron on the Bloch sphere). The infidelity is shown on the left-hand axis, defined as $1-F$ where $F$ is the fidelity (Eq.(\ref{eq:fidelity})) between the density matrix predicted by the model and that of the exact evolution of the spin-boson two-level system. For the 4-dimensional models (shown in blue) we also plot in red the largest real eigenvalue of the dynamical matrix of the model. The small this value the more stable the model. This value is shown on the right-hand axis.  
    \label{fig:dmd_4-5_vs_pos_eigen_vals}}
    See Ref. \cite{popovych_lsid_plot_2024} regarding a python code used to generate this figure.
\end{figure}

\subsection{Applying linear system identification to the spin-boson model}

We first apply the DMD method to the state vector formed from the block vector
and the trace of the density matrix. Our state vector is thus 
\begin{align} \label{eq:augmented_Bloch4}
\mathbf{y} = \begin{pmatrix} 
2 \operatorname{Re}(\rho_{01})\\
2 \operatorname{Im}(\rho_{10}) \\
\rho_{00} - \rho_{11} \\
\rho_{00} + \rho_{11} 
\end{pmatrix} 
\end{align}
and this gives us a 4-dimensional model for all values of $Q$. The fidelity of these models is shown using violin plots (light blue) in 
Fig.~\ref{fig:dmd_4-5_vs_pos_eigen_vals}. We see that this method obtains very accurate models for the highest four values of $Q$. We also plot the largest value of the real eigenvalues of the dynamical matrix for the models. This shows that for the highest $Q$ values, the models are also very stable. 

Quite striking is the fact that for the lowest three values of $Q$ the DMD method performs poorly; it does not even do as well as the best-fit Markovian master equations determined above, which also preserve complete positivity etc., and thus come with more constraints (compare with Fig.\ref{fig:kossak_vs_lindblad_fids}). 

We now attempt to use these 4-dimensional DMD models as the first stage in a process of fitting a master equation model --- that is, attempt to find the closest master equation to these models. First, comparing with the general form of the GKSL master equation, we find that the DMD models are not in this class. Clearly, there is no point in attempting to fit the models for the lowest three $Q$ values because the LSID model itself is not as good at the GKSL model obtained using polynomial optimization. For the lowest four $Q$ values we used a method similar to the one proposed by Zhang and Sarovar \cite{zhang_identification_2015} to fit master-equation models. This produced models that were one to two orders of magnitude less accurate in terms of infidelity than the models identified with the polynomial optimization method from the initial time series.

We can conclude that LSID produces models that can be very accurate for sufficiently low damping rates, but they do not correspond particularly closely to Markovian master equations. They are also not particularly useful as a stepping-stone to obtaining Markovian master equation models. For strong damping, LSID methods perform poorly.

Julia code for 4th rank DMD implementation for the identification of a two-level spin-boson system is available on GitHub by this link~\footnote{\url{https://github.com/zpopovych/quantum-open-systems-polynomial-sid/blob/main/LSID/01_DMD_Bloch4_sb.ipynb}}.

We are not quite done yet with LSID. As discussed above, we can also use LSID to obtain models with a higher dimension than that of the density matrix to reproduce the more accurately evolution of the density matrix. The higher the dimension we use, the more accurate we can expect the model to be. Essentially, the extra dimensions are taking into account the non-Markovian nature of the evolution, as they can contain the memory that the bath has about the previous evolution of the system. 

We use the DMD method described above to obtain a 16-dimensional model, and we find that truncating this model to five dimensions is sufficient to obtain accurate models of the spin-boson system for the entire range of $Q$-factors we explored. All of these models are also stable. While our focus here has been to develop a method to find models in which physical properties such as complete positivity and trace preservation are built in, this result shows that if the purpose is merely to obtain an accurate model of a non-Markovian system (over a long but nevertheless limited duration) LSID can do this very well by using models with increased dimension. Caution should be used with these models, however, when running them for significantly longer than the time series used to obtain them, since their future behavior is less certain than that in which physical constraints are built in.

Julia code for implementing the 5th-rank DMD and ERA methods for identification of the spin-boson system is available on GitHub~\footnote{\url{https://github.com/zpopovych/quantum-open-systems-polynomial-sid/blob/main/LSID/01_DMDvsERA_rank5_sb_trn4_tst20.ipynb}}.



\section{CONCLUSIONS \label{sec:summary}}

Here we have shown how methods that find the global optima of polynomial functions can be used to obtain the most accurate Markovian models for open quantum systems given time-series data of their evolution. This provides an efficient way to obtain the best Markovian models that are guaranteed to preserve the trace norm and the complete positivity of the density matrix, and can provide insights due to their physical interpretability. In addition, the ability to find the best possible Markovian model for a non-Markovian evolution provides an operational measure of non-Markovianity. 

We have shown that the method we introduce here is a significant advance over previous methods for obtaining Markovian master equation models as obtains much more accurate models. We have also explored the use of standard linear system identification methods (the Dynamical Mode Decomposition Algorithm and the Eigensystem Realization Algorithm) for obtaining models directly from time-series data. While these come without the guarantees of preserving the positivity of the density matrix, they are able to obtain models that have more degrees of freedom than the system density matrix and thus to directly model non-Markovian effects. By testing these methods on the spin-boson system we find that when using them to obtain a model with the same number of degrees of freedom as the density matrix (a Marovian model) they performed very well for sufficiently week damping (high $Q$) but surprisingly performed significantly worse than our polynomial optimization method for strong damping. Conversely, when allowed to use an additional degree of freedom, they obtained accurate models for all $Q$-factors that we explored.   

In applying both our polynomial optimization method and standard linear system identification  to the spin-boson system, we found a number of surprises. Among these were i) that the non-Markovianity of the system is not at all monotonic with increasing damping rate, ii) the lowest damping rate we explored was not the one that had the most accurate Markovian master equation model, and iii) Breuer's measure of non-Markovianity did not predict which damping rates would have the most accurate Markovian models. We also found that for very strong damping the spin-boson system can be described quite accurately with a Markovian model. 

While finding the best Markovian models for general open quantum extension is useful, it is likely even more useful to be able find even more accurate non-Markovian models for general open quantum systems that preserve complete positivity. We expect that the method we have introduced here can be readily extended to the identification of non-Markovian models, and this is an important direction for future work.

\bigskip
\emph{Acknowledgments:}
D.I.B. and Z.P. are supported by Army Research Office (ARO) (grant W911NF-23-1-0288, program manager Dr.~James Joseph) and by the DEVCOM Army Research Laboratory under cooperative agreement W911NF-21-2-0139.  G.K. and J.M. have been supported by 
the
Czech Science Foundation (23-07947S). This work has been partially supported by project MIS 5154714 of the National Recovery and Resilience Plan Greece 2.0 funded by the European Union under the NextGenerationEU Program.

\emph{Disclaimer (USG)} 
The views and conclusions contained in this document are those of the authors and should not be interpreted as representing the official policies, either expressed or implied, of ARO or the U.S. Government. The U.S. Government is authorized to reproduce and distribute reprints for Government purposes notwithstanding any copyright notation herein.

\emph{Disclaimer (HSBC)}.
This paper was prepared for information purposes
and is not a product of HSBC Bank Plc. or its affiliates.
Neither HSBC Bank Plc. nor any of its affiliates make
any explicit or implied representation or warranty and
none of them accept any liability in connection with
this paper, including, but not limited to, the completeness,
accuracy, reliability of information contained herein and
the potential legal, compliance, tax or accounting effects
thereof. Copyright HSBC Group 2024.

Z.P. and K.J. contributed equally to this work.

\appendix 

\section{Breuer's measure of non-Markovianity}
\label{AppA} 

Breuer's measure of the non-Markovianity of an evoltuion is defined by \cite{breuer_measure_2009}  
\begin{align}
\mathcal{N} = \max_{\substack{\rho^{(1)}(0) \\ \rho^{(2)}(0)}} \, \left[ \int_{\sigma>0}\!\!\!\!\!\sigma \,dt \right] 
\end{align}
where $\rho^{(1)}(0)$ and $\rho^{(2)}(0)$ all possible pairs of the initial states of the system and, $\sigma = \frac{dD}{dt}$ - is the time derivative of the trace distance $D$ between the corresponding points of the trajctories $\rho^{(1)}(t)$ and $\rho^{(2)}(t)$. The trace distance between trajectories at a particular time $t$ can be calculated as follows: 
\begin{align}
D(\rho^{(1)}, \rho^{(2)}, t) = \frac{1}{2} \operatorname{Tr}|\rho^{(1)}(t) - \rho^{(2)}(t)|,
\end{align}
in which $|A| = \sqrt{A^\dagger A}$.
As we had just ten effectively distinct evolutions of the spin-boson system we estimated the Breuer measure as follows:
\begin{align} \label{eq:nonmark_breuer}
\mathcal{N} = \max_{\substack{\rho^{(1)} \\ \rho^{(2)}}} \left[ \sum_{\Delta D_j >0}\! \Delta D_j \right] ,
\end{align}
where $\Delta D_j = D(\rho^{(1)}, \rho^{(2)}, t_j) - D(\rho^{(1)}, \rho^{(2)}, t_{j-1})$

Julia code for the calculations of non-Markovianity is available on GitHub at this link~\footnote{\url{https://github.com/zpopovych/quantum-open-systems-polynomial-sid/blob/main/NonMarkovianity.ipynb}}. 

\section{Linear system identification: mapping the model representation to the original state vector}
\label{AppB} 
In general the original system that generates the time series data has a state vector $\mathbf{v}(t)$ and the actual dynamical system we are interested in may have a smaller vector $\mathbf{y}(t) = \mathcal{C}\mathbf{v}(t)$. In our case for quantum system identification we usually think of $\mathcal{C}$ as being the identity (although if we think of the state vector $\mathbf{v}(t)$ as including the bath, then $\mathcal{C}$ is not the identity). In any case, it is only the time series of the density matrix of the two-level system (the state $\mathbf{y}(t)$) that provides the data for the system identification. 

The system identification method will in general obtain a dynamical system with state vector $\mathbf{x}(t)$ and this will determine $\mathbf{y}(t)$ via $\mathbf{y}(t) = K\mathbf{x}(t)$ for some constant matrix $K$. We want to find the invertible matrix $S$ such that 
\begin{align}
\mathbf{x}(t) = S \mathbf{v}(t) . 
\end{align}
To do this first define matrices $R$ and $Y$ formed by sets of state vectors by 
\begin{align}
    R = \left( \mathbf{v}_1 \cdots \mathbf{v}_{N} \right) , \;\;\; Y = \left( \mathbf{y}_1 \cdots \mathbf{y}_{N} \right) . 
\end{align}
Now we have 
\begin{align}
    Y = \mathcal{C} \left( \mathbf{x}_1 \cdots \mathbf{x}_{N} \right) = \mathcal{C}SR 
\end{align}
Now recall that if we denote the act of ``vectorizing" a matrix $Z$ (i.e., stacking all its columns one below the other to form a single column vector) as $\mbox{vec}[Z]$ then for any matrices $A$, $B$, $C$, we have $\mbox{vec}[ABC] = (C^{\ms{T}} \otimes A) \mbox{vec}[B]$. Applying this relation to both sides of the above equation we have 
\begin{align}
    \mbox{vec}[Y] = (R^{\ms{T}} \otimes \mathcal{C}) \mbox{vec}[S] 
\end{align}
or equivalently,
\begin{align}
    \mbox{vec}[S] = \left( R^{\ms{T}} \otimes \mathcal{C} \right)^{\ms{pinv}} \mbox{vec}[Y] 
\end{align}
in which ``pinv" stands for the Moore-Penrose pseudo-inverse.

\bibliography{references, refs_2, repositories}

\begin{thebibliography}{77}%
\makeatletter
\providecommand \@ifxundefined [1]{%
 \@ifx{#1\undefined}
}%
\providecommand \@ifnum [1]{%
 \ifnum #1\expandafter \@firstoftwo
 \else \expandafter \@secondoftwo
 \fi
}%
\providecommand \@ifx [1]{%
 \ifx #1\expandafter \@firstoftwo
 \else \expandafter \@secondoftwo
 \fi
}%
\providecommand \natexlab [1]{#1}%
\providecommand \enquote  [1]{``#1''}%
\providecommand \bibnamefont  [1]{#1}%
\providecommand \bibfnamefont [1]{#1}%
\providecommand \citenamefont [1]{#1}%
\providecommand \href@noop [0]{\@secondoftwo}%
\providecommand \href [0]{\begingroup \@sanitize@url \@href}%
\providecommand \@href[1]{\@@startlink{#1}\@@href}%
\providecommand \@@href[1]{\endgroup#1\@@endlink}%
\providecommand \@sanitize@url [0]{\catcode `\\12\catcode `\$12\catcode `\&12\catcode `\#12\catcode `\^12\catcode `\_12\catcode `\%12\relax}%
\providecommand \@@startlink[1]{}%
\providecommand \@@endlink[0]{}%
\providecommand \url  [0]{\begingroup\@sanitize@url \@url }%
\providecommand \@url [1]{\endgroup\@href {#1}{\urlprefix }}%
\providecommand \urlprefix  [0]{URL }%
\providecommand \Eprint [0]{\href }%
\providecommand \doibase [0]{http://dx.doi.org/}%
\providecommand \selectlanguage [0]{\@gobble}%
\providecommand \bibinfo  [0]{\@secondoftwo}%
\providecommand \bibfield  [0]{\@secondoftwo}%
\providecommand \translation [1]{[#1]}%
\providecommand \BibitemOpen [0]{}%
\providecommand \bibitemStop [0]{}%
\providecommand \bibitemNoStop [0]{.\EOS\space}%
\providecommand \EOS [0]{\spacefactor3000\relax}%
\providecommand \BibitemShut  [1]{\csname bibitem#1\endcsname}%
\let\auto@bib@innerbib\@empty
\bibitem [{\citenamefont {Nielsen}\ and\ \citenamefont {Chuang}(2010)}]{nielsen_quantum_2010}%
  \BibitemOpen
  \bibfield  {author} {\bibinfo {author} {\bibfnamefont {M.~A.}\ \bibnamefont {Nielsen}}\ and\ \bibinfo {author} {\bibfnamefont {I.~L.}\ \bibnamefont {Chuang}},\ }\href@noop {} {\emph {\bibinfo {title} {Quantum computation and quantum information}}},\ \bibinfo {edition} {10th}\ ed.\ (\bibinfo  {publisher} {Cambridge University Press},\ \bibinfo {address} {Cambridge ; New York},\ \bibinfo {year} {2010})\BibitemShut {NoStop}%
\bibitem [{\citenamefont {O'Brien}\ \emph {et~al.}(2009)\citenamefont {O'Brien}, \citenamefont {Furusawa},\ and\ \citenamefont {Vučković}}]{obrien_photonic_2009}%
  \BibitemOpen
  \bibfield  {author} {\bibinfo {author} {\bibfnamefont {J.~L.}\ \bibnamefont {O'Brien}}, \bibinfo {author} {\bibfnamefont {A.}~\bibnamefont {Furusawa}}, \ and\ \bibinfo {author} {\bibfnamefont {J.}~\bibnamefont {Vučković}},\ }\href {\doibase 10.1038/nphoton.2009.229} {\bibfield  {journal} {\bibinfo  {journal} {Nature Photonics}\ }\textbf {\bibinfo {volume} {3}},\ \bibinfo {pages} {687} (\bibinfo {year} {2009})}\BibitemShut {NoStop}%
\bibitem [{\citenamefont {Huggins}\ \emph {et~al.}(2022)\citenamefont {Huggins}, \citenamefont {O'Gorman}, \citenamefont {Rubin}, \citenamefont {Reichman}, \citenamefont {Babbush},\ and\ \citenamefont {Lee}}]{huggins_unbiasing_2022}%
  \BibitemOpen
  \bibfield  {author} {\bibinfo {author} {\bibfnamefont {W.~J.}\ \bibnamefont {Huggins}}, \bibinfo {author} {\bibfnamefont {B.~A.}\ \bibnamefont {O'Gorman}}, \bibinfo {author} {\bibfnamefont {N.~C.}\ \bibnamefont {Rubin}}, \bibinfo {author} {\bibfnamefont {D.~R.}\ \bibnamefont {Reichman}}, \bibinfo {author} {\bibfnamefont {R.}~\bibnamefont {Babbush}}, \ and\ \bibinfo {author} {\bibfnamefont {J.}~\bibnamefont {Lee}},\ }\href {\doibase 10.1038/s41586-021-04351-z} {\bibfield  {journal} {\bibinfo  {journal} {Nature}\ }\textbf {\bibinfo {volume} {603}},\ \bibinfo {pages} {416} (\bibinfo {year} {2022})}\BibitemShut {NoStop}%
\bibitem [{\citenamefont {Yu}\ \emph {et~al.}(2021)\citenamefont {Yu}, \citenamefont {Wang}, \citenamefont {Dong},\ and\ \citenamefont {Petersen}}]{yu_capability_2021}%
  \BibitemOpen
  \bibfield  {author} {\bibinfo {author} {\bibfnamefont {Q.}~\bibnamefont {Yu}}, \bibinfo {author} {\bibfnamefont {Y.}~\bibnamefont {Wang}}, \bibinfo {author} {\bibfnamefont {D.}~\bibnamefont {Dong}}, \ and\ \bibinfo {author} {\bibfnamefont {I.~R.}\ \bibnamefont {Petersen}},\ }\href {\doibase 10.1016/j.automatica.2021.109612} {\bibfield  {journal} {\bibinfo  {journal} {Automatica}\ }\textbf {\bibinfo {volume} {129}},\ \bibinfo {pages} {109612} (\bibinfo {year} {2021})}\BibitemShut {NoStop}%
\bibitem [{\citenamefont {Xie}\ \emph {et~al.}(2022)\citenamefont {Xie}, \citenamefont {Yu}, \citenamefont {Rodgers}, \citenamefont {Xu}, \citenamefont {Chi-Durán}, \citenamefont {Toros}, \citenamefont {Quack}, \citenamefont {De~Leon},\ and\ \citenamefont {Maurer}}]{xie_biocompatible_2022}%
  \BibitemOpen
  \bibfield  {author} {\bibinfo {author} {\bibfnamefont {M.}~\bibnamefont {Xie}}, \bibinfo {author} {\bibfnamefont {X.}~\bibnamefont {Yu}}, \bibinfo {author} {\bibfnamefont {L.~V.~H.}\ \bibnamefont {Rodgers}}, \bibinfo {author} {\bibfnamefont {D.}~\bibnamefont {Xu}}, \bibinfo {author} {\bibfnamefont {I.}~\bibnamefont {Chi-Durán}}, \bibinfo {author} {\bibfnamefont {A.}~\bibnamefont {Toros}}, \bibinfo {author} {\bibfnamefont {N.}~\bibnamefont {Quack}}, \bibinfo {author} {\bibfnamefont {N.~P.}\ \bibnamefont {De~Leon}}, \ and\ \bibinfo {author} {\bibfnamefont {P.~C.}\ \bibnamefont {Maurer}},\ }\href {\doibase 10.1073/pnas.2114186119} {\bibfield  {journal} {\bibinfo  {journal} {Proceedings of the National Academy of Sciences}\ }\textbf {\bibinfo {volume} {119}},\ \bibinfo {pages} {e2114186119} (\bibinfo {year} {2022})}\BibitemShut {NoStop}%
\bibitem [{\citenamefont {Marciniak}\ \emph {et~al.}(2022)\citenamefont {Marciniak}, \citenamefont {Feldker}, \citenamefont {Pogorelov}, \citenamefont {Kaubruegger}, \citenamefont {Vasilyev}, \citenamefont {Van~Bijnen}, \citenamefont {Schindler}, \citenamefont {Zoller}, \citenamefont {Blatt},\ and\ \citenamefont {Monz}}]{marciniak_optimal_2022}%
  \BibitemOpen
  \bibfield  {author} {\bibinfo {author} {\bibfnamefont {C.~D.}\ \bibnamefont {Marciniak}}, \bibinfo {author} {\bibfnamefont {T.}~\bibnamefont {Feldker}}, \bibinfo {author} {\bibfnamefont {I.}~\bibnamefont {Pogorelov}}, \bibinfo {author} {\bibfnamefont {R.}~\bibnamefont {Kaubruegger}}, \bibinfo {author} {\bibfnamefont {D.~V.}\ \bibnamefont {Vasilyev}}, \bibinfo {author} {\bibfnamefont {R.}~\bibnamefont {Van~Bijnen}}, \bibinfo {author} {\bibfnamefont {P.}~\bibnamefont {Schindler}}, \bibinfo {author} {\bibfnamefont {P.}~\bibnamefont {Zoller}}, \bibinfo {author} {\bibfnamefont {R.}~\bibnamefont {Blatt}}, \ and\ \bibinfo {author} {\bibfnamefont {T.}~\bibnamefont {Monz}},\ }\href {\doibase 10.1038/s41586-022-04435-4} {\bibfield  {journal} {\bibinfo  {journal} {Nature}\ }\textbf {\bibinfo {volume} {603}},\ \bibinfo {pages} {604} (\bibinfo {year} {2022})}\BibitemShut {NoStop}%
\bibitem [{\citenamefont {Ljung}(1987)}]{ljung_system_1987}%
  \BibitemOpen
  \bibfield  {author} {\bibinfo {author} {\bibfnamefont {L.}~\bibnamefont {Ljung}},\ }\href@noop {} {\emph {\bibinfo {title} {System identification: theory for the user}}},\ Prentice-{Hall} information and system sciences series\ (\bibinfo  {publisher} {Prentice-Hall},\ \bibinfo {address} {Englewood Cliffs, NJ},\ \bibinfo {year} {1987})\BibitemShut {NoStop}%
\bibitem [{\citenamefont {McCauley}\ \emph {et~al.}(2020)\citenamefont {McCauley}, \citenamefont {Cruikshank}, \citenamefont {Bondar},\ and\ \citenamefont {Jacobs}}]{mccauley_accurate_2020}%
  \BibitemOpen
  \bibfield  {author} {\bibinfo {author} {\bibfnamefont {G.}~\bibnamefont {McCauley}}, \bibinfo {author} {\bibfnamefont {B.}~\bibnamefont {Cruikshank}}, \bibinfo {author} {\bibfnamefont {D.~I.}\ \bibnamefont {Bondar}}, \ and\ \bibinfo {author} {\bibfnamefont {K.}~\bibnamefont {Jacobs}},\ }\href {\doibase 10.1038/s41534-020-00299-6} {\bibfield  {journal} {\bibinfo  {journal} {npj Quantum Information}\ }\textbf {\bibinfo {volume} {6}},\ \bibinfo {pages} {1} (\bibinfo {year} {2020})}\BibitemShut {NoStop}%
\bibitem [{\citenamefont {Zhang}\ and\ \citenamefont {Sarovar}(2015)}]{zhang_identification_2015}%
  \BibitemOpen
  \bibfield  {author} {\bibinfo {author} {\bibfnamefont {J.}~\bibnamefont {Zhang}}\ and\ \bibinfo {author} {\bibfnamefont {M.}~\bibnamefont {Sarovar}},\ }\href {\doibase 10.1103/PhysRevA.91.052121} {\bibfield  {journal} {\bibinfo  {journal} {Physical Review A}\ }\textbf {\bibinfo {volume} {91}},\ \bibinfo {pages} {052121} (\bibinfo {year} {2015})}\BibitemShut {NoStop}%
\bibitem [{\citenamefont {Samach}\ \emph {et~al.}(2022)\citenamefont {Samach}, \citenamefont {Greene}, \citenamefont {Borregaard}, \citenamefont {Christandl}, \citenamefont {Barreto}, \citenamefont {Kim}, \citenamefont {McNally}, \citenamefont {Melville}, \citenamefont {Niedzielski}, \citenamefont {Sung}, \citenamefont {Rosenberg}, \citenamefont {Schwartz}, \citenamefont {Yoder}, \citenamefont {Orlando}, \citenamefont {Wang}, \citenamefont {Gustavsson}, \citenamefont {Kjaergaard},\ and\ \citenamefont {Oliver}}]{samach_lindblad_2022}%
  \BibitemOpen
  \bibfield  {author} {\bibinfo {author} {\bibfnamefont {G.~O.}\ \bibnamefont {Samach}}, \bibinfo {author} {\bibfnamefont {A.}~\bibnamefont {Greene}}, \bibinfo {author} {\bibfnamefont {J.}~\bibnamefont {Borregaard}}, \bibinfo {author} {\bibfnamefont {M.}~\bibnamefont {Christandl}}, \bibinfo {author} {\bibfnamefont {J.}~\bibnamefont {Barreto}}, \bibinfo {author} {\bibfnamefont {D.~K.}\ \bibnamefont {Kim}}, \bibinfo {author} {\bibfnamefont {C.~M.}\ \bibnamefont {McNally}}, \bibinfo {author} {\bibfnamefont {A.}~\bibnamefont {Melville}}, \bibinfo {author} {\bibfnamefont {B.~M.}\ \bibnamefont {Niedzielski}}, \bibinfo {author} {\bibfnamefont {Y.}~\bibnamefont {Sung}}, \bibinfo {author} {\bibfnamefont {D.}~\bibnamefont {Rosenberg}}, \bibinfo {author} {\bibfnamefont {M.~E.}\ \bibnamefont {Schwartz}}, \bibinfo {author} {\bibfnamefont {J.~L.}\ \bibnamefont {Yoder}}, \bibinfo {author} {\bibfnamefont {T.~P.}\ \bibnamefont {Orlando}}, \bibinfo {author} {\bibfnamefont {J.~I.-J.}\ \bibnamefont {Wang}}, \bibinfo {author}
  {\bibfnamefont {S.}~\bibnamefont {Gustavsson}}, \bibinfo {author} {\bibfnamefont {M.}~\bibnamefont {Kjaergaard}}, \ and\ \bibinfo {author} {\bibfnamefont {W.~D.}\ \bibnamefont {Oliver}},\ }\href {\doibase 10.1103/PhysRevApplied.18.064056} {\bibfield  {journal} {\bibinfo  {journal} {Physical Review Applied}\ }\textbf {\bibinfo {volume} {18}},\ \bibinfo {pages} {064056} (\bibinfo {year} {2022})}\BibitemShut {NoStop}%
\bibitem [{\citenamefont {Xue}\ \emph {et~al.}(2021)\citenamefont {Xue}, \citenamefont {Wu}, \citenamefont {Ma}, \citenamefont {Li},\ and\ \citenamefont {Jiang}}]{xue_gradient_2021}%
  \BibitemOpen
  \bibfield  {author} {\bibinfo {author} {\bibfnamefont {S.}~\bibnamefont {Xue}}, \bibinfo {author} {\bibfnamefont {R.}~\bibnamefont {Wu}}, \bibinfo {author} {\bibfnamefont {S.}~\bibnamefont {Ma}}, \bibinfo {author} {\bibfnamefont {D.}~\bibnamefont {Li}}, \ and\ \bibinfo {author} {\bibfnamefont {M.}~\bibnamefont {Jiang}},\ }\href {\doibase 10.1103/PhysRevA.103.022604} {\bibfield  {journal} {\bibinfo  {journal} {Physical Review A}\ }\textbf {\bibinfo {volume} {103}},\ \bibinfo {pages} {022604} (\bibinfo {year} {2021})}\BibitemShut {NoStop}%
\bibitem [{\citenamefont {Mazza}\ \emph {et~al.}(2021)\citenamefont {Mazza}, \citenamefont {Zietlow}, \citenamefont {Carollo}, \citenamefont {Andergassen}, \citenamefont {Martius},\ and\ \citenamefont {Lesanovsky}}]{mazza_machine_2021}%
  \BibitemOpen
  \bibfield  {author} {\bibinfo {author} {\bibfnamefont {P.~P.}\ \bibnamefont {Mazza}}, \bibinfo {author} {\bibfnamefont {D.}~\bibnamefont {Zietlow}}, \bibinfo {author} {\bibfnamefont {F.}~\bibnamefont {Carollo}}, \bibinfo {author} {\bibfnamefont {S.}~\bibnamefont {Andergassen}}, \bibinfo {author} {\bibfnamefont {G.}~\bibnamefont {Martius}}, \ and\ \bibinfo {author} {\bibfnamefont {I.}~\bibnamefont {Lesanovsky}},\ }\href {\doibase 10.1103/PhysRevResearch.3.023084} {\bibfield  {journal} {\bibinfo  {journal} {Physical Review Research}\ }\textbf {\bibinfo {volume} {3}},\ \bibinfo {pages} {023084} (\bibinfo {year} {2021})}\BibitemShut {NoStop}%
\bibitem [{\citenamefont {Henrion}\ \emph {et~al.}(2021)\citenamefont {Henrion}, \citenamefont {Korda},\ and\ \citenamefont {Lasserre}}]{henrion_moment-sos_2021}%
  \BibitemOpen
  \bibfield  {author} {\bibinfo {author} {\bibfnamefont {D.}~\bibnamefont {Henrion}}, \bibinfo {author} {\bibfnamefont {M.}~\bibnamefont {Korda}}, \ and\ \bibinfo {author} {\bibfnamefont {J.-B.}\ \bibnamefont {Lasserre}},\ }\href@noop {} {\emph {\bibinfo {title} {The moment-{SOS} hierarchy: lectures in probability, statistics, computational geometry, control and nonlinear {PDEs}}}},\ \bibinfo {series} {Series on optimization and its applications}\ No.\ \bibinfo {number} {vol. 4}\ (\bibinfo  {publisher} {World Scientific},\ \bibinfo {address} {New Jersey},\ \bibinfo {year} {2021})\BibitemShut {NoStop}%
\bibitem [{\citenamefont {Parrilo}(2000)}]{parrilo2000structured}%
  \BibitemOpen
  \bibfield  {author} {\bibinfo {author} {\bibfnamefont {P.~A.}\ \bibnamefont {Parrilo}},\ }\href@noop {} {\emph {\bibinfo {title} {Structured semidefinite programs and semialgebraic geometry methods in robustness and optimization}}}\ (\bibinfo  {publisher} {California Institute of Technology},\ \bibinfo {year} {2000})\BibitemShut {NoStop}%
\bibitem [{\citenamefont {Lasserre}(2001)}]{lasserre_global_2001}%
  \BibitemOpen
  \bibfield  {author} {\bibinfo {author} {\bibfnamefont {J.~B.}\ \bibnamefont {Lasserre}},\ }\href {\doibase 10.1137/S1052623400366802} {\bibfield  {journal} {\bibinfo  {journal} {SIAM Journal on Optimization}\ }\textbf {\bibinfo {volume} {11}},\ \bibinfo {pages} {796} (\bibinfo {year} {2001})}\BibitemShut {NoStop}%
\bibitem [{\citenamefont {Nie}(2023)}]{nie_moment_2023}%
  \BibitemOpen
  \bibfield  {author} {\bibinfo {author} {\bibfnamefont {J.}~\bibnamefont {Nie}},\ }\href@noop {} {\emph {\bibinfo {title} {Moment and polynomial optimization}}},\ {MOS}-{SIAM} series on optimization\ (\bibinfo  {publisher} {Society for Industrial and Applied Mathematics},\ \bibinfo {address} {Philadelphia},\ \bibinfo {year} {2023})\BibitemShut {NoStop}%
\bibitem [{\citenamefont {Wang}\ \emph {et~al.}(2019)\citenamefont {Wang}, \citenamefont {Magron},\ and\ \citenamefont {Lasserre}}]{wang2019tssos}%
  \BibitemOpen
  \bibfield  {author} {\bibinfo {author} {\bibfnamefont {J.}~\bibnamefont {Wang}}, \bibinfo {author} {\bibfnamefont {V.}~\bibnamefont {Magron}}, \ and\ \bibinfo {author} {\bibfnamefont {J.-B.}\ \bibnamefont {Lasserre}},\ }\href@noop {} {\bibfield  {journal} {\bibinfo  {journal} {arXiv preprint arXiv:1912.08899}\ } (\bibinfo {year} {2019})}\BibitemShut {NoStop}%
\bibitem [{\citenamefont {Wang}\ \emph {et~al.}(2020)\citenamefont {Wang}, \citenamefont {Magron},\ and\ \citenamefont {Lasserre}}]{wang_tssos_2020}%
  \BibitemOpen
  \bibfield  {author} {\bibinfo {author} {\bibfnamefont {J.}~\bibnamefont {Wang}}, \bibinfo {author} {\bibfnamefont {V.}~\bibnamefont {Magron}}, \ and\ \bibinfo {author} {\bibfnamefont {J.-B.}\ \bibnamefont {Lasserre}},\ }\href {http://arxiv.org/abs/1912.08899} {\enquote {\bibinfo {title} {{TSSOS}: {A} {Moment}-{SOS} hierarchy that exploits term sparsity},}\ } (\bibinfo {year} {2020}),\ \bibinfo {note} {arXiv:1912.08899 [math]}\BibitemShut {NoStop}%
\bibitem [{\citenamefont {Lofberg}(2004)}]{lofberg2004yalmip}%
  \BibitemOpen
  \bibfield  {author} {\bibinfo {author} {\bibfnamefont {J.}~\bibnamefont {Lofberg}},\ }in\ \href@noop {} {\emph {\bibinfo {booktitle} {2004 IEEE international conference on robotics and automation (IEEE Cat. No. 04CH37508)}}}\ (\bibinfo {organization} {IEEE},\ \bibinfo {year} {2004})\ pp.\ \bibinfo {pages} {284--289}\BibitemShut {NoStop}%
\bibitem [{\citenamefont {Heller}\ and\ \citenamefont {Pajdla}(2016)}]{heller2016gposolver}%
  \BibitemOpen
  \bibfield  {author} {\bibinfo {author} {\bibfnamefont {J.}~\bibnamefont {Heller}}\ and\ \bibinfo {author} {\bibfnamefont {T.}~\bibnamefont {Pajdla}},\ }\href@noop {} {\bibfield  {journal} {\bibinfo  {journal} {Optimization Methods and Software}\ }\textbf {\bibinfo {volume} {31}},\ \bibinfo {pages} {405} (\bibinfo {year} {2016})}\BibitemShut {NoStop}%
\bibitem [{\citenamefont {Bulla}\ \emph {et~al.}(2003)\citenamefont {Bulla}, \citenamefont {Tong},\ and\ \citenamefont {Vojta}}]{bulla_numerical_2003}%
  \BibitemOpen
  \bibfield  {author} {\bibinfo {author} {\bibfnamefont {R.}~\bibnamefont {Bulla}}, \bibinfo {author} {\bibfnamefont {N.-H.}\ \bibnamefont {Tong}}, \ and\ \bibinfo {author} {\bibfnamefont {M.}~\bibnamefont {Vojta}},\ }\href {\doibase 10.1103/PhysRevLett.91.170601} {\bibfield  {journal} {\bibinfo  {journal} {Physical Review Letters}\ }\textbf {\bibinfo {volume} {91}},\ \bibinfo {pages} {170601} (\bibinfo {year} {2003})}\BibitemShut {NoStop}%
\bibitem [{\citenamefont {Chin}\ \emph {et~al.}(2010)\citenamefont {Chin}, \citenamefont {Rivas}, \citenamefont {Huelga},\ and\ \citenamefont {Plenio}}]{chin_exact_2010}%
  \BibitemOpen
  \bibfield  {author} {\bibinfo {author} {\bibfnamefont {A.~W.}\ \bibnamefont {Chin}}, \bibinfo {author} {\bibfnamefont {Ã.}~\bibnamefont {Rivas}}, \bibinfo {author} {\bibfnamefont {S.~F.}\ \bibnamefont {Huelga}}, \ and\ \bibinfo {author} {\bibfnamefont {M.~B.}\ \bibnamefont {Plenio}},\ }\href {\doibase 10.1063/1.3490188} {\bibfield  {journal} {\bibinfo  {journal} {Journal of Mathematical Physics}\ }\textbf {\bibinfo {volume} {51}},\ \bibinfo {pages} {092109} (\bibinfo {year} {2010})}\BibitemShut {NoStop}%
\bibitem [{\citenamefont {Prior}\ \emph {et~al.}(2010)\citenamefont {Prior}, \citenamefont {Chin}, \citenamefont {Huelga},\ and\ \citenamefont {Plenio}}]{prior_efficient_2010}%
  \BibitemOpen
  \bibfield  {author} {\bibinfo {author} {\bibfnamefont {J.}~\bibnamefont {Prior}}, \bibinfo {author} {\bibfnamefont {A.~W.}\ \bibnamefont {Chin}}, \bibinfo {author} {\bibfnamefont {S.~F.}\ \bibnamefont {Huelga}}, \ and\ \bibinfo {author} {\bibfnamefont {M.~B.}\ \bibnamefont {Plenio}},\ }\href {\doibase 10.1103/PhysRevLett.105.050404} {\bibfield  {journal} {\bibinfo  {journal} {Physical Review Letters}\ }\textbf {\bibinfo {volume} {105}},\ \bibinfo {pages} {050404} (\bibinfo {year} {2010})}\BibitemShut {NoStop}%
\bibitem [{\citenamefont {Kutz}(2016)}]{kutz_dynamic_2016}%
  \BibitemOpen
  \bibinfo {editor} {\bibfnamefont {J.~N.}\ \bibnamefont {Kutz}},\ ed.,\ \href@noop {} {\emph {\bibinfo {title} {Dynamic mode decomposition: data-driven modeling of complex systems}}}\ (\bibinfo  {publisher} {Society for Industrial and Applied Mathematics},\ \bibinfo {address} {Philadelphia},\ \bibinfo {year} {2016})\BibitemShut {NoStop}%
\bibitem [{\citenamefont {Goldschmidt}(2022)}]{goldschmidt_data-driven_2022}%
  \BibitemOpen
  \bibfield  {author} {\bibinfo {author} {\bibfnamefont {A.~J.}\ \bibnamefont {Goldschmidt}},\ }\emph {\bibinfo {title} {Data-driven modeling and control of quantum dynamics}},\ \href {https://digital.lib.washington.edu/researchworks/bitstream/handle/1773/49426/Goldschmidt_washington_0250E_24609.pdf?sequence=1&isAllowed=y} {Ph.D. thesis},\ \bibinfo  {school} {University of Washington}, \bibinfo {address} {Seattle, WA 98195} (\bibinfo {year} {2022})\BibitemShut {NoStop}%
\bibitem [{\citenamefont {Gorini}\ \emph {et~al.}(1976)\citenamefont {Gorini}, \citenamefont {Kossakowski},\ and\ \citenamefont {Sudarshan}}]{gorini_completely_1976}%
  \BibitemOpen
  \bibfield  {author} {\bibinfo {author} {\bibfnamefont {V.}~\bibnamefont {Gorini}}, \bibinfo {author} {\bibfnamefont {A.}~\bibnamefont {Kossakowski}}, \ and\ \bibinfo {author} {\bibfnamefont {E.~C.~G.}\ \bibnamefont {Sudarshan}},\ }\href {\doibase 10.1063/1.522979} {\bibfield  {journal} {\bibinfo  {journal} {Journal of Mathematical Physics}\ }\textbf {\bibinfo {volume} {17}},\ \bibinfo {pages} {821} (\bibinfo {year} {1976})}\BibitemShut {NoStop}%
\bibitem [{\citenamefont {Lindblad}(1976)}]{lindblad_generators_1976}%
  \BibitemOpen
  \bibfield  {author} {\bibinfo {author} {\bibfnamefont {G.}~\bibnamefont {Lindblad}},\ }\href {\doibase 10.1007/BF01608499} {\bibfield  {journal} {\bibinfo  {journal} {Communications in Mathematical Physics}\ }\textbf {\bibinfo {volume} {48}},\ \bibinfo {pages} {119} (\bibinfo {year} {1976})}\BibitemShut {NoStop}%
\bibitem [{\citenamefont {Li}\ \emph {et~al.}(2018{\natexlab{a}})\citenamefont {Li}, \citenamefont {Hall},\ and\ \citenamefont {Wiseman}}]{li_concepts_2018}%
  \BibitemOpen
  \bibfield  {author} {\bibinfo {author} {\bibfnamefont {L.}~\bibnamefont {Li}}, \bibinfo {author} {\bibfnamefont {M.~J.~W.}\ \bibnamefont {Hall}}, \ and\ \bibinfo {author} {\bibfnamefont {H.~M.}\ \bibnamefont {Wiseman}},\ }\href {\doibase 10.1016/j.physrep.2018.07.001} {\bibfield  {journal} {\bibinfo  {journal} {Physics Reports}\ }\textbf {\bibinfo {volume} {759}},\ \bibinfo {pages} {1} (\bibinfo {year} {2018}{\natexlab{a}})},\ \bibinfo {note} {arXiv:1712.08879 [quant-ph]}\BibitemShut {NoStop}%
\bibitem [{\citenamefont {Howard}\ \emph {et~al.}(2006)\citenamefont {Howard}, \citenamefont {Twamley}, \citenamefont {Wittmann}, \citenamefont {Gaebel}, \citenamefont {Jelezko},\ and\ \citenamefont {Wrachtrup}}]{howard_quantum_2006}%
  \BibitemOpen
  \bibfield  {author} {\bibinfo {author} {\bibfnamefont {M.}~\bibnamefont {Howard}}, \bibinfo {author} {\bibfnamefont {J.}~\bibnamefont {Twamley}}, \bibinfo {author} {\bibfnamefont {C.}~\bibnamefont {Wittmann}}, \bibinfo {author} {\bibfnamefont {T.}~\bibnamefont {Gaebel}}, \bibinfo {author} {\bibfnamefont {F.}~\bibnamefont {Jelezko}}, \ and\ \bibinfo {author} {\bibfnamefont {J.}~\bibnamefont {Wrachtrup}},\ }\href {\doibase 10.1088/1367-2630/8/3/033} {\bibfield  {journal} {\bibinfo  {journal} {New Journal of Physics}\ }\textbf {\bibinfo {volume} {8}},\ \bibinfo {pages} {33} (\bibinfo {year} {2006})}\BibitemShut {NoStop}%
\bibitem [{\citenamefont {Zhang}\ \emph {et~al.}(2021)\citenamefont {Zhang}, \citenamefont {Wei}, \citenamefont {Yan}, \citenamefont {Man}, \citenamefont {Xia},\ and\ \citenamefont {Fan}}]{zhang_non-markovian_2021}%
  \BibitemOpen
  \bibfield  {author} {\bibinfo {author} {\bibfnamefont {Y.-J.}\ \bibnamefont {Zhang}}, \bibinfo {author} {\bibfnamefont {H.}~\bibnamefont {Wei}}, \bibinfo {author} {\bibfnamefont {W.-B.}\ \bibnamefont {Yan}}, \bibinfo {author} {\bibfnamefont {Z.-X.}\ \bibnamefont {Man}}, \bibinfo {author} {\bibfnamefont {Y.-J.}\ \bibnamefont {Xia}}, \ and\ \bibinfo {author} {\bibfnamefont {H.}~\bibnamefont {Fan}},\ }\href {\doibase 10.1088/1367-2630/ac2c2a} {\bibfield  {journal} {\bibinfo  {journal} {New Journal of Physics}\ }\textbf {\bibinfo {volume} {23}},\ \bibinfo {pages} {113004} (\bibinfo {year} {2021})}\BibitemShut {NoStop}%
\bibitem [{\citenamefont {Ben~Av}\ \emph {et~al.}(2020)\citenamefont {Ben~Av}, \citenamefont {Shapira}, \citenamefont {Akerman},\ and\ \citenamefont {Ozeri}}]{ben_av_direct_2020}%
  \BibitemOpen
  \bibfield  {author} {\bibinfo {author} {\bibfnamefont {E.}~\bibnamefont {Ben~Av}}, \bibinfo {author} {\bibfnamefont {Y.}~\bibnamefont {Shapira}}, \bibinfo {author} {\bibfnamefont {N.}~\bibnamefont {Akerman}}, \ and\ \bibinfo {author} {\bibfnamefont {R.}~\bibnamefont {Ozeri}},\ }\href {\doibase 10.1103/PhysRevA.101.062305} {\bibfield  {journal} {\bibinfo  {journal} {Physical Review A}\ }\textbf {\bibinfo {volume} {101}},\ \bibinfo {pages} {062305} (\bibinfo {year} {2020})}\BibitemShut {NoStop}%
\bibitem [{\citenamefont {Dobrynin}\ \emph {et~al.}(2024)\citenamefont {Dobrynin}, \citenamefont {Cardarelli}, \citenamefont {Müller},\ and\ \citenamefont {Bermudez}}]{dobrynin_compressed-sensing_2024}%
  \BibitemOpen
  \bibfield  {author} {\bibinfo {author} {\bibfnamefont {D.}~\bibnamefont {Dobrynin}}, \bibinfo {author} {\bibfnamefont {L.}~\bibnamefont {Cardarelli}}, \bibinfo {author} {\bibfnamefont {M.}~\bibnamefont {Müller}}, \ and\ \bibinfo {author} {\bibfnamefont {A.}~\bibnamefont {Bermudez}},\ }\href {\doibase 10.48550/ARXIV.2403.07462} {\enquote {\bibinfo {title} {Compressed-sensing {Lindbladian} quantum tomography with trapped ions},}\ } (\bibinfo {year} {2024}),\ \bibinfo {note} {version Number: 1}\BibitemShut {NoStop}%
\bibitem [{\citenamefont {Jacobs}(2014)}]{jacobs_quantum_2014}%
  \BibitemOpen
  \bibfield  {author} {\bibinfo {author} {\bibfnamefont {K.}~\bibnamefont {Jacobs}},\ }\href@noop {} {\emph {\bibinfo {title} {Quantum measurement theory and its applications}}}\ (\bibinfo  {publisher} {Cambridge University Press},\ \bibinfo {address} {Cambridge},\ \bibinfo {year} {2014})\BibitemShut {NoStop}%
\bibitem [{\citenamefont {Yoshida}(2024)}]{yoshida2024uniqueness}%
  \BibitemOpen
  \bibfield  {author} {\bibinfo {author} {\bibfnamefont {H.}~\bibnamefont {Yoshida}},\ }\href@noop {} {\bibfield  {journal} {\bibinfo  {journal} {Physical Review A}\ }\textbf {\bibinfo {volume} {109}},\ \bibinfo {pages} {022218} (\bibinfo {year} {2024})}\BibitemShut {NoStop}%
\bibitem [{\citenamefont {Henrion}\ and\ \citenamefont {Lasserre}(2011)}]{henrion2011inner}%
  \BibitemOpen
  \bibfield  {author} {\bibinfo {author} {\bibfnamefont {D.}~\bibnamefont {Henrion}}\ and\ \bibinfo {author} {\bibfnamefont {J.-B.}\ \bibnamefont {Lasserre}},\ }\href@noop {} {\bibfield  {journal} {\bibinfo  {journal} {IEEE Transactions on Automatic Control}\ }\textbf {\bibinfo {volume} {57}},\ \bibinfo {pages} {1456} (\bibinfo {year} {2011})}\BibitemShut {NoStop}%
\bibitem [{\citenamefont {Kocvara}\ and\ \citenamefont {Stingl}(2012)}]{kocvara2012pennon}%
  \BibitemOpen
  \bibfield  {author} {\bibinfo {author} {\bibfnamefont {M.}~\bibnamefont {Kocvara}}\ and\ \bibinfo {author} {\bibfnamefont {M.}~\bibnamefont {Stingl}},\ }\href@noop {} {\bibfield  {journal} {\bibinfo  {journal} {Handbook on semidefinite, conic and polynomial optimization}\ ,\ \bibinfo {pages} {755}} (\bibinfo {year} {2012})}\BibitemShut {NoStop}%
\bibitem [{\citenamefont {Zheng}\ and\ \citenamefont {Fantuzzi}(2023)}]{zheng2023sum}%
  \BibitemOpen
  \bibfield  {author} {\bibinfo {author} {\bibfnamefont {Y.}~\bibnamefont {Zheng}}\ and\ \bibinfo {author} {\bibfnamefont {G.}~\bibnamefont {Fantuzzi}},\ }\href@noop {} {\bibfield  {journal} {\bibinfo  {journal} {Mathematical Programming}\ }\textbf {\bibinfo {volume} {197}},\ \bibinfo {pages} {71} (\bibinfo {year} {2023})}\BibitemShut {NoStop}%
\bibitem [{\citenamefont {Guo}\ and\ \citenamefont {Wang}(2024)}]{guo2024moment}%
  \BibitemOpen
  \bibfield  {author} {\bibinfo {author} {\bibfnamefont {F.}~\bibnamefont {Guo}}\ and\ \bibinfo {author} {\bibfnamefont {J.}~\bibnamefont {Wang}},\ }\href@noop {} {\bibfield  {journal} {\bibinfo  {journal} {Mathematics of Operations Research}\ } (\bibinfo {year} {2024})}\BibitemShut {NoStop}%
\bibitem [{\citenamefont {Leggett}\ \emph {et~al.}(1987)\citenamefont {Leggett}, \citenamefont {Chakravarty}, \citenamefont {Dorsey}, \citenamefont {Fisher}, \citenamefont {Garg},\ and\ \citenamefont {Zwerger}}]{leggett_dynamics_1987}%
  \BibitemOpen
  \bibfield  {author} {\bibinfo {author} {\bibfnamefont {A.~J.}\ \bibnamefont {Leggett}}, \bibinfo {author} {\bibfnamefont {S.}~\bibnamefont {Chakravarty}}, \bibinfo {author} {\bibfnamefont {A.~T.}\ \bibnamefont {Dorsey}}, \bibinfo {author} {\bibfnamefont {M.~P.~A.}\ \bibnamefont {Fisher}}, \bibinfo {author} {\bibfnamefont {A.}~\bibnamefont {Garg}}, \ and\ \bibinfo {author} {\bibfnamefont {W.}~\bibnamefont {Zwerger}},\ }\href {\doibase 10.1103/RevModPhys.59.1} {\bibfield  {journal} {\bibinfo  {journal} {Reviews of Modern Physics}\ }\textbf {\bibinfo {volume} {59}},\ \bibinfo {pages} {1} (\bibinfo {year} {1987})}\BibitemShut {NoStop}%
\bibitem [{\citenamefont {Santra}\ \emph {et~al.}(2017)\citenamefont {Santra}, \citenamefont {Cruikshank}, \citenamefont {Balu},\ and\ \citenamefont {Jacobs}}]{santra_fermis_2017}%
  \BibitemOpen
  \bibfield  {author} {\bibinfo {author} {\bibfnamefont {S.}~\bibnamefont {Santra}}, \bibinfo {author} {\bibfnamefont {B.}~\bibnamefont {Cruikshank}}, \bibinfo {author} {\bibfnamefont {R.}~\bibnamefont {Balu}}, \ and\ \bibinfo {author} {\bibfnamefont {K.}~\bibnamefont {Jacobs}},\ }\href {\doibase 10.1088/1751-8121/aa8777} {\bibfield  {journal} {\bibinfo  {journal} {Journal of Physics A: Mathematical and Theoretical}\ }\textbf {\bibinfo {volume} {50}},\ \bibinfo {pages} {415302} (\bibinfo {year} {2017})}\BibitemShut {NoStop}%
\bibitem [{\citenamefont {Vidal}(2003)}]{vidal_efficient_2003}%
  \BibitemOpen
  \bibfield  {author} {\bibinfo {author} {\bibfnamefont {G.}~\bibnamefont {Vidal}},\ }\href {\doibase 10.1103/PhysRevLett.91.147902} {\bibfield  {journal} {\bibinfo  {journal} {Physical Review Letters}\ }\textbf {\bibinfo {volume} {91}},\ \bibinfo {pages} {147902} (\bibinfo {year} {2003})}\BibitemShut {NoStop}%
\bibitem [{\citenamefont {Vidal}(2004)}]{vidal_efficient_2004}%
  \BibitemOpen
  \bibfield  {author} {\bibinfo {author} {\bibfnamefont {G.}~\bibnamefont {Vidal}},\ }\href {\doibase 10.1103/PhysRevLett.93.040502} {\bibfield  {journal} {\bibinfo  {journal} {Physical Review Letters}\ }\textbf {\bibinfo {volume} {93}},\ \bibinfo {pages} {040502} (\bibinfo {year} {2004})}\BibitemShut {NoStop}%
\bibitem [{\citenamefont {Davies}(1974)}]{Davies1974}%
  \BibitemOpen
  \bibfield  {author} {\bibinfo {author} {\bibfnamefont {E.~B.}\ \bibnamefont {Davies}},\ }\href {http://projecteuclid.org/euclid.cmp/1103860160} {\bibfield  {journal} {\bibinfo  {journal} {Comm. Math. Phys.}\ }\textbf {\bibinfo {volume} {39}},\ \bibinfo {pages} {91} (\bibinfo {year} {1974})}\BibitemShut {NoStop}%
\bibitem [{\citenamefont {Gnutzmann}\ and\ \citenamefont {Haake}(1996)}]{Gnutzmann96}%
  \BibitemOpen
  \bibfield  {author} {\bibinfo {author} {\bibfnamefont {S.}~\bibnamefont {Gnutzmann}}\ and\ \bibinfo {author} {\bibfnamefont {F.}~\bibnamefont {Haake}},\ }\href@noop {} {\bibfield  {journal} {\bibinfo  {journal} {Z. Phys. B}\ }\textbf {\bibinfo {volume} {101}},\ \bibinfo {pages} {263} (\bibinfo {year} {1996})}\BibitemShut {NoStop}%
\bibitem [{\citenamefont {Gaspard}\ and\ \citenamefont {Nagaoka}(1999)}]{Gaspard99}%
  \BibitemOpen
  \bibfield  {author} {\bibinfo {author} {\bibfnamefont {P.}~\bibnamefont {Gaspard}}\ and\ \bibinfo {author} {\bibfnamefont {M.}~\bibnamefont {Nagaoka}},\ }\href@noop {} {\bibfield  {journal} {\bibinfo  {journal} {J. Chem. Phys.}\ }\textbf {\bibinfo {volume} {111}},\ \bibinfo {pages} {5668} (\bibinfo {year} {1999})}\BibitemShut {NoStop}%
\bibitem [{\citenamefont {Hu}\ \emph {et~al.}(1992)\citenamefont {Hu}, \citenamefont {Paz},\ and\ \citenamefont {Zhang}}]{Hu92}%
  \BibitemOpen
  \bibfield  {author} {\bibinfo {author} {\bibfnamefont {B.~L.}\ \bibnamefont {Hu}}, \bibinfo {author} {\bibfnamefont {J.~P.}\ \bibnamefont {Paz}}, \ and\ \bibinfo {author} {\bibfnamefont {Y.}~\bibnamefont {Zhang}},\ }\href@noop {} {\bibfield  {journal} {\bibinfo  {journal} {Phys. Rev. D}\ }\textbf {\bibinfo {volume} {45}},\ \bibinfo {pages} {2843} (\bibinfo {year} {1992})}\BibitemShut {NoStop}%
\bibitem [{\citenamefont {Kossakowski}(1973)}]{kossakowski_general_1973}%
  \BibitemOpen
  \bibfield  {author} {\bibinfo {author} {\bibfnamefont {A.}~\bibnamefont {Kossakowski}},\ }\href@noop {} {\bibfield  {journal} {\bibinfo  {journal} {Bulletin de l'Academie Polonaise des Sciences. Serie des Sciences, Mathematiques, Astronomiques et Physiques}\ }\textbf {\bibinfo {volume} {21}},\ \bibinfo {pages} {649} (\bibinfo {year} {1973})}\BibitemShut {NoStop}%
\bibitem [{\citenamefont {Jozsa}(1994)}]{jozsa_fidelity_1994}%
  \BibitemOpen
  \bibfield  {author} {\bibinfo {author} {\bibfnamefont {R.}~\bibnamefont {Jozsa}},\ }\href {\doibase 10.1080/09500349414552171} {\bibfield  {journal} {\bibinfo  {journal} {Journal of Modern Optics}\ }\textbf {\bibinfo {volume} {41}},\ \bibinfo {pages} {2315} (\bibinfo {year} {1994})}\BibitemShut {NoStop}%
\bibitem [{\citenamefont {Popovych}(2024{\natexlab{a}})}]{popovych_kossak_pop_2024}%
  \BibitemOpen
  \bibfield  {author} {\bibinfo {author} {\bibfnamefont {Z.}~\bibnamefont {Popovych}},\ }\href {https://github.com/zpopovych/quantum-open-systems-polynomial-sid/blob/main/01_Kossakowski_CONSTR_train-test.ipynb} {\enquote {\bibinfo {title} {Identification of kossakowski form markovian master equation},}\ } (\bibinfo {year} {2024}{\natexlab{a}})\BibitemShut {NoStop}%
\bibitem [{\citenamefont {Popovych}(2024{\natexlab{b}})}]{popovych_lindblad_pop_2024}%
  \BibitemOpen
  \bibfield  {author} {\bibinfo {author} {\bibfnamefont {Z.}~\bibnamefont {Popovych}},\ }\href {https://github.com/zpopovych/quantum-open-systems-polynomial-sid/blob/main/01_Lindblad_CONSTR_train-test.ipynb} {\enquote {\bibinfo {title} {Identification of lindblad form markovian master equation},}\ } (\bibinfo {year} {2024}{\natexlab{b}})\BibitemShut {NoStop}%
\bibitem [{\citenamefont {Popovych}(2024{\natexlab{c}})}]{popovych_kossak_benchmark_2024}%
  \BibitemOpen
  \bibfield  {author} {\bibinfo {author} {\bibfnamefont {Z.}~\bibnamefont {Popovych}},\ }\href {https://github.com/zpopovych/quantum-open-systems-polynomial-sid/blob/main/01_Benchmark_Kossak_NLOPT_SLSQP.ipynb} {\enquote {\bibinfo {title} {Identification of kossakowski benchmark form markovian master equation},}\ } (\bibinfo {year} {2024}{\natexlab{c}})\BibitemShut {NoStop}%
\bibitem [{\citenamefont {Popovych}(2024{\natexlab{d}})}]{popovych_lindblad_benchmark_2024}%
  \BibitemOpen
  \bibfield  {author} {\bibinfo {author} {\bibfnamefont {Z.}~\bibnamefont {Popovych}},\ }\href {https://github.com/zpopovych/quantum-open-systems-polynomial-sid/blob/main/01_Benchmark_Lindblad_NLOPT_SLSQP.ipynb} {\enquote {\bibinfo {title} {Identification of lindblad benchmark form markovian master equation},}\ } (\bibinfo {year} {2024}{\natexlab{d}})\BibitemShut {NoStop}%
\bibitem [{\citenamefont {Breuer}\ \emph {et~al.}(2009)\citenamefont {Breuer}, \citenamefont {Laine},\ and\ \citenamefont {Piilo}}]{breuer_measure_2009}%
  \BibitemOpen
  \bibfield  {author} {\bibinfo {author} {\bibfnamefont {H.-P.}\ \bibnamefont {Breuer}}, \bibinfo {author} {\bibfnamefont {E.-M.}\ \bibnamefont {Laine}}, \ and\ \bibinfo {author} {\bibfnamefont {J.}~\bibnamefont {Piilo}},\ }\href {\doibase 10.1103/PhysRevLett.103.210401} {\bibfield  {journal} {\bibinfo  {journal} {Physical Review Letters}\ }\textbf {\bibinfo {volume} {103}},\ \bibinfo {pages} {210401} (\bibinfo {year} {2009})}\BibitemShut {NoStop}%
\bibitem [{\citenamefont {Popovych}(2024{\natexlab{e}})}]{popovych_non-mark-estims_2024}%
  \BibitemOpen
  \bibfield  {author} {\bibinfo {author} {\bibfnamefont {Z.}~\bibnamefont {Popovych}},\ }\href {https://github.com/zpopovych/quantum-open-systems-polynomial-sid/blob/main/NonMarkovianity.ipynb} {\enquote {\bibinfo {title} {Non-markovianity estimations},}\ } (\bibinfo {year} {2024}{\natexlab{e}})\BibitemShut {NoStop}%
\bibitem [{\citenamefont {Popovych}(2024{\natexlab{f}})}]{popovych_pop_plot_2024}%
  \BibitemOpen
  \bibfield  {author} {\bibinfo {author} {\bibfnamefont {Z.}~\bibnamefont {Popovych}},\ }\href {https://github.com/zpopovych/quantum-open-systems-polynomial-sid/blob/main/03_plot_violin.ipynb} {\enquote {\bibinfo {title} {Plotting fidelity of pop methods},}\ } (\bibinfo {year} {2024}{\natexlab{f}})\BibitemShut {NoStop}%
\bibitem [{\citenamefont {Popovych}(2024{\natexlab{g}})}]{popovych_pop_plot_by_train_2024}%
  \BibitemOpen
  \bibfield  {author} {\bibinfo {author} {\bibfnamefont {Z.}~\bibnamefont {Popovych}},\ }\href {https://github.com/zpopovych/quantum-open-systems-polynomial-sid/blob/main/06_plot_Kossak_and_Lindblad_by_duration.ipynb} {\enquote {\bibinfo {title} {Plot pop sid by duatation of training},}\ } (\bibinfo {year} {2024}{\natexlab{g}})\BibitemShut {NoStop}%
\bibitem [{\citenamefont {Johnson}(2007)}]{steven_g_johnson_nlopt_2007}%
  \BibitemOpen
  \bibfield  {author} {\bibinfo {author} {\bibfnamefont {S.~G.}\ \bibnamefont {Johnson}},\ }\href {https://github.com/stevengj/nlopt} {\enquote {\bibinfo {title} {The {NLopt} nonlinear-optimization package},}\ } (\bibinfo {year} {2007})\BibitemShut {NoStop}%
\bibitem [{\citenamefont {Boggs}\ and\ \citenamefont {Tolle}(1995)}]{boggs1995sequential}%
  \BibitemOpen
  \bibfield  {author} {\bibinfo {author} {\bibfnamefont {P.~T.}\ \bibnamefont {Boggs}}\ and\ \bibinfo {author} {\bibfnamefont {J.~W.}\ \bibnamefont {Tolle}},\ }\href@noop {} {\bibfield  {journal} {\bibinfo  {journal} {Acta numerica}\ }\textbf {\bibinfo {volume} {4}},\ \bibinfo {pages} {1} (\bibinfo {year} {1995})}\BibitemShut {NoStop}%
\bibitem [{\citenamefont {Gill}\ and\ \citenamefont {Wong}(2011)}]{gill2011sequential}%
  \BibitemOpen
  \bibfield  {author} {\bibinfo {author} {\bibfnamefont {P.~E.}\ \bibnamefont {Gill}}\ and\ \bibinfo {author} {\bibfnamefont {E.}~\bibnamefont {Wong}},\ }in\ \href@noop {} {\emph {\bibinfo {booktitle} {Mixed integer nonlinear programming}}}\ (\bibinfo  {publisher} {Springer},\ \bibinfo {year} {2011})\ pp.\ \bibinfo {pages} {147--224}\BibitemShut {NoStop}%
\bibitem [{Note1()}]{Note1}%
  \BibitemOpen
  \bibinfo {note} {\protect \url {https://github.com/zpopovych/quantum-open-systems-polynomial-sid/blob/main/01_Kossakowski_CONSTR_train-test.ipynb}}\BibitemShut {NoStop}%
\bibitem [{Note2()}]{Note2}%
  \BibitemOpen
  \bibinfo {note} {\protect \url {https://github.com/zpopovych/quantum-open-systems-polynomial-sid/blob/main/KOSSAK_CONSTR_TSSOS_treshold_1e-15_FROB_QO_2024-Sep-06_at_11-57.h5} (Note: The file consists of the groups that correspond to the damping rates $\gamma $. In each group you can find the Kossakowski matrix in dataset \protect \texttt {"C"} and Hamiltonian in dataset \protect \texttt {"H"})}\BibitemShut {NoStop}%
\bibitem [{\citenamefont {Li}\ \emph {et~al.}(2018{\natexlab{b}})\citenamefont {Li}, \citenamefont {Hall},\ and\ \citenamefont {Wiseman}}]{Li_2018}%
  \BibitemOpen
  \bibfield  {author} {\bibinfo {author} {\bibfnamefont {L.}~\bibnamefont {Li}}, \bibinfo {author} {\bibfnamefont {M.~J.}\ \bibnamefont {Hall}}, \ and\ \bibinfo {author} {\bibfnamefont {H.~M.}\ \bibnamefont {Wiseman}},\ }\href {\doibase https://doi.org/10.1016/j.physrep.2018.07.001} {\bibfield  {journal} {\bibinfo  {journal} {Physics Reports}\ }\textbf {\bibinfo {volume} {759}},\ \bibinfo {pages} {1} (\bibinfo {year} {2018}{\natexlab{b}})}\BibitemShut {NoStop}%
\bibitem [{Note3()}]{Note3}%
  \BibitemOpen
  \bibinfo {note} {\protect \url {https://github.com/zpopovych/quantum-open-systems-polynomial-sid/blob/main/01_Lindblad_CONSTR_train-test.ipynb}}\BibitemShut {NoStop}%
\bibitem [{Note4()}]{Note4}%
  \BibitemOpen
  \bibinfo {note} {\protect \url {https://github.com/zpopovych/quantum-open-systems-polynomial-sid/blob/main/LINDBLAD4_CONSTR_TSSOS_treshold_1e-9_FROB_QO_2024-Sep-06_at_16-19.h5}}\BibitemShut {NoStop}%
\bibitem [{\citenamefont {Mabuchi}(1996)}]{mabuchi_dynamical_1996}%
  \BibitemOpen
  \bibfield  {author} {\bibinfo {author} {\bibfnamefont {H.}~\bibnamefont {Mabuchi}},\ }\href {\doibase 10.1088/1355-5111/8/6/002} {\bibfield  {journal} {\bibinfo  {journal} {Quantum and Semiclassical Optics: Journal of the European Optical Society Part B}\ }\textbf {\bibinfo {volume} {8}},\ \bibinfo {pages} {1103} (\bibinfo {year} {1996})}\BibitemShut {NoStop}%
\bibitem [{\citenamefont {Budišić}\ \emph {et~al.}(2012)\citenamefont {Budišić}, \citenamefont {Mohr},\ and\ \citenamefont {Mezić}}]{budisic_applied_2012}%
  \BibitemOpen
  \bibfield  {author} {\bibinfo {author} {\bibfnamefont {M.}~\bibnamefont {Budišić}}, \bibinfo {author} {\bibfnamefont {R.}~\bibnamefont {Mohr}}, \ and\ \bibinfo {author} {\bibfnamefont {I.}~\bibnamefont {Mezić}},\ }\href {\doibase 10.1063/1.4772195} {\bibfield  {journal} {\bibinfo  {journal} {Chaos: An Interdisciplinary Journal of Nonlinear Science}\ }\textbf {\bibinfo {volume} {22}},\ \bibinfo {pages} {047510} (\bibinfo {year} {2012})}\BibitemShut {NoStop}%
\bibitem [{\citenamefont {Zhang}\ and\ \citenamefont {Sarovar}(2014)}]{zhang_quantum_2014}%
  \BibitemOpen
  \bibfield  {author} {\bibinfo {author} {\bibfnamefont {J.}~\bibnamefont {Zhang}}\ and\ \bibinfo {author} {\bibfnamefont {M.}~\bibnamefont {Sarovar}},\ }\href {\doibase 10.1103/PhysRevLett.113.080401} {\bibfield  {journal} {\bibinfo  {journal} {Physical Review Letters}\ }\textbf {\bibinfo {volume} {113}},\ \bibinfo {pages} {080401} (\bibinfo {year} {2014})}\BibitemShut {NoStop}%
\bibitem [{\citenamefont {Juang}\ and\ \citenamefont {Pappa}(1985{\natexlab{a}})}]{juang_eigensystem_1985}%
  \BibitemOpen
  \bibfield  {author} {\bibinfo {author} {\bibfnamefont {J.-N.}\ \bibnamefont {Juang}}\ and\ \bibinfo {author} {\bibfnamefont {R.~S.}\ \bibnamefont {Pappa}},\ }\href {\doibase 10.2514/3.20031} {\bibfield  {journal} {\bibinfo  {journal} {Journal of Guidance, Control, and Dynamics}\ }\textbf {\bibinfo {volume} {8}},\ \bibinfo {pages} {620} (\bibinfo {year} {1985}{\natexlab{a}})}\BibitemShut {NoStop}%
\bibitem [{\citenamefont {Juang}(1994)}]{juang_applied_1994}%
  \BibitemOpen
  \bibfield  {author} {\bibinfo {author} {\bibfnamefont {J.-N.}\ \bibnamefont {Juang}},\ }\href@noop {} {\emph {\bibinfo {title} {Applied system identification}}}\ (\bibinfo  {publisher} {Prentice Hall},\ \bibinfo {address} {Englewood Cliffs, N.J},\ \bibinfo {year} {1994})\BibitemShut {NoStop}%
\bibitem [{\citenamefont {Baddoo}\ \emph {et~al.}(2023)\citenamefont {Baddoo}, \citenamefont {Herrmann}, \citenamefont {McKeon}, \citenamefont {Nathan~Kutz},\ and\ \citenamefont {Brunton}}]{baddoo_physics-informed_2023}%
  \BibitemOpen
  \bibfield  {author} {\bibinfo {author} {\bibfnamefont {P.~J.}\ \bibnamefont {Baddoo}}, \bibinfo {author} {\bibfnamefont {B.}~\bibnamefont {Herrmann}}, \bibinfo {author} {\bibfnamefont {B.~J.}\ \bibnamefont {McKeon}}, \bibinfo {author} {\bibfnamefont {J.}~\bibnamefont {Nathan~Kutz}}, \ and\ \bibinfo {author} {\bibfnamefont {S.~L.}\ \bibnamefont {Brunton}},\ }\href {\doibase 10.1098/rspa.2022.0576} {\bibfield  {journal} {\bibinfo  {journal} {Proceedings of the Royal Society A: Mathematical, Physical and Engineering Sciences}\ }\textbf {\bibinfo {volume} {479}},\ \bibinfo {pages} {20220576} (\bibinfo {year} {2023})}\BibitemShut {NoStop}%
\bibitem [{\citenamefont {Schmid}(2010)}]{SCHMID_2010}%
  \BibitemOpen
  \bibfield  {author} {\bibinfo {author} {\bibfnamefont {P.~J.}\ \bibnamefont {Schmid}},\ }\href {\doibase 10.1017/S0022112010001217} {\bibfield  {journal} {\bibinfo  {journal} {Journal of Fluid Mechanics}\ }\textbf {\bibinfo {volume} {656}},\ \bibinfo {pages} {5–28} (\bibinfo {year} {2010})}\BibitemShut {NoStop}%
\bibitem [{\citenamefont {Juang}\ and\ \citenamefont {Pappa}(1985{\natexlab{b}})}]{Juang85}%
  \BibitemOpen
  \bibfield  {author} {\bibinfo {author} {\bibfnamefont {J.-N.}\ \bibnamefont {Juang}}\ and\ \bibinfo {author} {\bibfnamefont {R.~S.}\ \bibnamefont {Pappa}},\ }\href {\doibase 10.2514/3.20031} {\bibfield  {journal} {\bibinfo  {journal} {Journal of Guidance, Control, and Dynamics}\ }\textbf {\bibinfo {volume} {8}},\ \bibinfo {pages} {620} (\bibinfo {year} {1985}{\natexlab{b}})}\BibitemShut {NoStop}%
\bibitem [{\citenamefont {Katayama}(2005)}]{Katayama05}%
  \BibitemOpen
  \bibfield  {author} {\bibinfo {author} {\bibfnamefont {T.}~\bibnamefont {Katayama}},\ }\href@noop {} {\emph {\bibinfo {title} {Subspace Methods for System Identification}}}\ (\bibinfo  {publisher} {Springer},\ \bibinfo {address} {New York},\ \bibinfo {year} {2005})\BibitemShut {NoStop}%
\bibitem [{\citenamefont {Popovych}(2024{\natexlab{h}})}]{popovych_lsid_plot_2024}%
  \BibitemOpen
  \bibfield  {author} {\bibinfo {author} {\bibfnamefont {Z.}~\bibnamefont {Popovych}},\ }\href {https://github.com/zpopovych/quantum-open-systems-polynomial-sid/blob/main/LSID/03_plot_LSID_violin.ipynb} {\enquote {\bibinfo {title} {Plot of linear sid performance},}\ } (\bibinfo {year} {2024}{\natexlab{h}})\BibitemShut {NoStop}%
\bibitem [{Note5()}]{Note5}%
  \BibitemOpen
  \bibinfo {note} {\protect \url {https://github.com/zpopovych/quantum-open-systems-polynomial-sid/blob/main/LSID/01_DMD_Bloch4_sb.ipynb}}\BibitemShut {NoStop}%
\bibitem [{Note6()}]{Note6}%
  \BibitemOpen
  \bibinfo {note} {\protect \url {https://github.com/zpopovych/quantum-open-systems-polynomial-sid/blob/main/LSID/01_DMDvsERA_rank5_sb_trn4_tst20.ipynb}}\BibitemShut {NoStop}%
\bibitem [{Note7()}]{Note7}%
  \BibitemOpen
  \bibinfo {note} {\protect \url {https://github.com/zpopovych/quantum-open-systems-polynomial-sid/blob/main/NonMarkovianity.ipynb}}\BibitemShut {NoStop}%
\end{thebibliography}%

\end{document}